\documentclass{aa}
\usepackage{graphicx}
\usepackage[utf8]{inputenc}
\usepackage{txfonts}
\usepackage{xspace}
\usepackage{microtype}
\usepackage{footmisc}
\usepackage[colorlinks=true,linkcolor=red,urlcolor=black,citecolor=blue]{hyperref}
\usepackage{lscape}

\newcommand{\tworxp}{2RXP~J130159.6$-$635806\xspace}
\newcommand{\aOF}{A0535$+$26\xspace}
\newcommand{\cep}{Cep~X-4\xspace}
\newcommand{\gxTOF}{GX~304$-$1\xspace}
\newcommand{\gro}{GRO~J1008$-$57\xspace}
\newcommand{\maxiZero}{MAXI~J0655$-$013\xspace}
\newcommand{\maxiOne}{MAXI~J1409$-$619\xspace}
\newcommand{\srga}{SRGA~J124404.1$-$632232\xspace}
\newcommand{\xper}{X~Per\xspace}
\newcommand{\igr}{IGR~J21347$+$4737\xspace}
\newcommand{\twoSXPS}{2SXPS~J075542.5$-$293353\xspace}
\newcommand{\sgr}{SGR~0755$-$2933\xspace}

\newcommand{\nustar}{\textit{NuSTAR}\xspace}
\newcommand{\swift}{\textit{Swift}\xspace}
\newcommand{\xrt}{\textit{Swift}/XRT\xspace}
\newcommand{\xmm}{\textit{XMM-Newton}\xspace}
\newcommand{\nicer}{\textit{NICER}\xspace}

\newcommand{\chandra}{\textit{Chandra}\xspace}
\newcommand{\integral}{\textit{INTEGRAL}\xspace}

\newcommand{\doublehump}{\texttt{doublehump}\xspace}
\newcommand{\polcap}{\texttt{polcap}\xspace}

\defcitealias{ballhausen2017}{Ba17}
\defcitealias{delgado-marti2001}{De01}
\defcitealias{doroshenko2021}{Do21}
\defcitealias{doroshenko2022}{Do22}
\defcitealias{finger1994}{Fi94}
\defcitealias{ghising2023a}{Gh23}
\defcitealias{gorban2022a}{Go22}
\defcitealias{kuhnel2013}{Ku13}
\defcitealias{lutovinov2021}{Lu21}
\defcitealias{malacaria2015}{Ma15}
\defcitealias{mcbride2007}{Mc07}
\defcitealias{mroz2021}{Mr21}
\defcitealias{pike2023}{Pi23}
\defcitealias{rai2023}{Ra23}
\defcitealias{rai2025}{Ra25}
\defcitealias{sugizaki2015}{Su15}
\defcitealias{vybornov2017}{Vy17}

\begin{document}

\title{
A simple relation: Neutron star magnetic field strength and spectral shape at low mass accretion rates}

\author{Nicolas~Zalot\inst{\ref{inst:remeis}}\thanks{\email{nicolas.zalot@fau.de}}
  \and
  Ekaterina~Sokolova-Lapa \inst{\ref{inst:remeis}} \and
  Aafia~Zainab \inst{\ref{inst:remeis}} \and
  Philipp~Thalhammer \inst{\ref{inst:remeis}} \and
  Jakob~Stierhof \inst{\ref{inst:remeis}} \and
  Katrin~Berger \inst{\ref{inst:remeis}} \and
  Katja~Pottschmidt\inst{\ref{inst:CRESST},\ref{inst:gsfc}}\thanks{deceased 17 June  2025}
  \and
  Ralf~Ballhausen \inst{\ref{inst:umd},\ref{inst:gsfc}} \and
  Christian~Malacaria\inst{\ref{inst:INAF-OAR}} \and
  Esin~Gulbahar\inst{\ref{inst:remeis}} \and
  J\"orn~Wilms \inst{\ref{inst:remeis}}
}

\institute{
  Dr.\ Karl-Remeis-Observatory and Erlangen Centre for Astroparticle Physics,
  Friedrich-Alexander-Universit\"at Erlangen-N\"urnberg,
  Sternwartstr.~7, 96049 Bamberg, Germany \label{inst:remeis}
  \and
  CRESST and Center for Space Sciences and Technology, University of Maryland, Baltimore County, 1000 Hilltop Circle, Baltimore, MD 21250, USA \label{inst:CRESST} \and 
  NASA Goddard Space Flight Center, Astrophysics Science Division, Greenbelt, MD 20771, USA \label{inst:gsfc} \and
  University of Maryland, Department of Astronomy, College Park, MD 20742, USA \label{inst:umd} \and
  INAF-Osservatorio Astronomico di Roma, Via Frascati 33, 00078, Monte Porzio Catone (RM), Italy \label{inst:INAF-OAR}
}

\date{Received December 19, 2025 / Accepted March 23, 2026}

\authorrunning{N. Zalot et al.}

\abstract
{The X-ray spectra of neutron stars with moderate magnetic fields ($B\sim 10^{12}$\,G) in high-mass X-ray binaries (HMXBs) at low X-ray luminosities ($L_\mathrm{X}\lesssim 10^{35}\,\mathrm{erg}\,\mathrm{s}^{-1}$) are characterized by a double humped shape. This shape has been explained either as the radiation from a two-temperature magnetized atmosphere, where thermal radiation dominates at soft X-rays below about 10\,keV, and cyclotron radiation with an imprinted cyclotron line dominates at high energies, or by the complex redistribution of primary X-rays in a structured atmosphere.}
{The theoretical explanations of the double humped structure predict the spectra to depend on the magnetic field. We aim to connect the model predictions with observations.}
{We analyzed archival \nustar observations of four HMXBs consisting of a neutron star and a Be~star (BeXRBs), with known magnetic fields at luminosities low enough to show the characteristic double-hump spectrum. We modeled these spectra empirically and derived a relation between the energy of the intersection of the two humps and the magnetic field strength. In a second step, we tested whether this correlation is supported by fitting synthetic spectra simulated with the physically self-consistent \polcap model.}
{We find a linear correlation between the magnetic field strength and the intersection energy for the real BeXRB \nustar spectra and \polcap-based simulated \nustar spectra alike.}
{The effect of the magnetic field on spectral formation results in an observable correlation between the field strength and spectral shape. This derived positive correlation between intersection energy and magnetic field strength also allowed us to roughly estimate the magnetic field strength via our proposed 2-B-12 rule. Additional observations of XRBs and dedicated modeling efforts will be necessary to determine whether this approach is valid beyond the $B$-field range of a few~$10^{12}$\,G to $10^{13}$\,G that was tested in this work.}


\maketitle
\nolinenumbers

\section{Introduction}\label{sec:introduction}
Magnetic fields play an important role in our understanding of the environment of magnetized neutron stars \citep[NSs; see, e.g.,][]{harding2006}.
Neutron stars are typically found in binary systems with either a low-mass companion \citep[low-mass X-ray binaries or LMXBs; see, e.g.,][]{mitsuda1984} or a high-mass companion \citep[high-mass X-ray binaries or HMXBs; see, e.g.,][]{chaty2011}.
In the latter case, the companion is typically a supergiant or Be-type massive star.
In HMXBs, mass accretion from the stellar companion onto the NS \citep[see, e.g.,][]{ghosh1979} can occur via accretion disks around the NS or via strong stellar winds from the companion.
The strong gravitational field of the neutron star governs the dynamics of the accreted matter at large distances.
At intermediate distances, however, the neutron star's strong magnetic field (or $B$-field) of $10^{12}$--$10^{13}$\,G affects the dynamics of the accreted material \citep[see, e.g.,][]{lamb1973}.
It modifies the radiative processes involved in reprocessing of radiation close to the neutron star surface and has a crucial effect on spectral formation \citep{basko1975,araya1999}.

Magnetic fields of strongly magnetized accreting neutron stars can often be measured directly via cyclotron resonant scattering features (CRSFs), or ``cyclotron lines,'' in their X-ray spectra \citep[see][for a review]{staubert2019}.
Cyclotron resonant scattering features are formed due to the quantization of electron momenta perpendicular to the magnetic field axis.
Compton scattering in optically thick media and the redistribution of photons with energies close to the energy difference between two adjacent Landau levels result in apparent absorption features with centroid energies close to this energy difference.
Cyclotron lines therefore probe the magnetic field strength within the line-forming region, which is typically assumed to be close to the neutron star surface.
The observed centroid energy of the CRSF, $E_\mathrm{CRSF}$, and the $B$-field strength are related via the 12-$B$-12 rule,
\begin{equation}
	E_\mathrm{CRSF}\sim \frac{n}{1+z} 11.6\,\mathrm{keV} \times \frac{B}{10^{12}\,\mathrm{G}},
	\label{eq:12B12}
\end{equation}
where $n$ corresponds to the $n^\mathrm{th}$ harmonic feature (i.e., $n=1$ corresponds to the fundamental line), $z$ is the gravitational redshift due to the NS gravitational field ($z=0.24$ for $M_\mathrm{NS}=1.4\,M_\odot$ and $R_\mathrm{NS}=12$\,km), and $B$ is the magnetic field strength in the line-forming region.

\begin{figure}
\centering
	\includegraphics[width=\columnwidth]{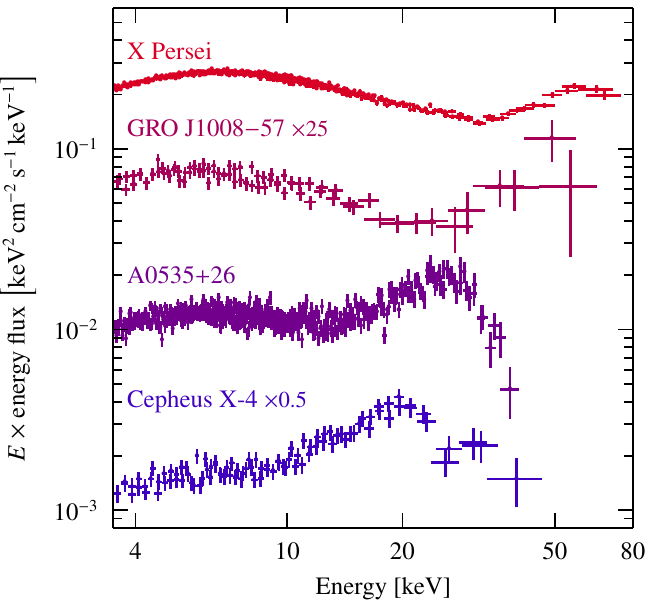}
	\caption{\nustar X-ray spectra of BeXRBs in quiescence. The spectra are ordered by decreasing energy at which the hard X-ray-component peaks, from top to bottom. 
	The respective rescaling factors, where required, are given next to the source names.}
	\label{fig:threesources}
\end{figure}

Cyclotron resonant scattering features, however,  are not the only spectral signatures that can arise from electron quantization.
At low mass accretion rates, ${\lesssim}10^{16}\,\mathrm{g}\,\mathrm{s}^{-1}$, when the pressure of the emitted radiation is insufficient to affect the deceleration of the accretion flow \citep[see, e.g.,][for a discussion of the transition to the radiation-dominated regime]{basko1976, becker2012, mushtukov2015a}, it is mainly decelerated by Coulomb collisions \citep[see][and references therein]{meszaros1983, harding1984, miller1987}.
The kinematics of the braking process in a strong magnetic field differs significantly from the non-magnetized case.
Coulomb collisions can result in multiple excitations of the ambient electrons \citep{miller1987,miller1989,nelson1993}, followed by radiative decay and the production of cyclotron photons.
This nonthermal (bulk) cyclotron emission, reprocessed in the atmosphere, is thought to produce a noticeable excess in the X-ray spectra \citep[e.g.,][]{nelson1995,mushtukov2021}.
The (Coulomb-heated) atmospheres consist of two distinct temperature zones, with significantly higher temperatures at the top than deep within the atmosphere and an outward decreasing electron density \citep[see, e.g.,][]{zeldovich1969, meszaros1983, miller1989}.
These effects are thought to be responsible for the formation of the two-component spectra observed in Be~X-ray binaries (BeXRBs) at typical luminosities of $10^{33}$--$10^{34}\,\mathrm{erg}\,\mathrm{s}^{-1}$ \citep[see, e.g.,][and Fig.~\ref{fig:threesources}]{tsygankov2019a,tsygankov2019,lutovinov2021}.
\citet{mushtukov2021} showed that the X-ray spectra of a two-temperature magnetized atmosphere -- with distinct cold and hot regions and cyclotron emission -- exhibit a characteristic double-hump shape: thermal emission dominates at soft X-rays, while reprocessed cyclotron emission, imprinted with a CRSF, appears at higher energies.
In their model, nearly all the accretion energy is assumed to be converted into cyclotron emission.
However, detailed kinematic calculations indicate that for magnetic fields of ${\sim} 5\times 10^{12}$\,G, the amount of electron excitations per falling proton can remain relatively low \citep{miller1989}, potentially reducing the nonthermal cyclotron contribution.
\citet{sokolova-lapa2021} explored an alternative scenario in which excitation plays only a minor role in spectral formation.
In this case, the structure of the atmosphere is affected by strong temperature and density gradients, polarization effects, and complex redistribution during the scattering process in the vicinity of the cyclotron resonance, which still results in a two-component X-ray spectrum with distinct low- and high-energy humps.
Both models, that of \citet{mushtukov2021} and that of \citet{sokolova-lapa2021}, have been successfully applied to describe the quiescent spectra of BeXRBs.
In both models, the formation of the high-energy component is closely related to cyclotron interactions.

In this paper we perform a systematic study of the X-ray spectra of BeXRBs at low luminosity. 
In Sect.~\ref{sec:observation}, we model X-ray spectra of a selection of BeXRBs with an empirical model consisting of two spectral humps and show that the location of these humps depends on the surface magnetic field of the neutron stars. In Sect.~\ref{sec:theory}, we then use theoretical spectral models for BeXRBs at low mass accretion rates to show that they can in principle reproduce the observed magnetic field dependency. In Sect.~\ref{sec:discussion}, we put our results in a physical context and discuss predictions by theoretical models. 

\section{X-ray spectral analysis of BeXRBs in quiescence}\label{sec:observation}
In this section we present our spectral analysis of nine BeXRBs.
We describe our source selection in Sect.~\ref{sec:sourceselection}
In Sect.~\ref{sec:obs:models}, we introduce semi-physical and phenomenological models that are typically utilized to describe the X-ray spectra of BeXRBs and present the empirical model utilized in this  work.
In Sect.~\ref{sec:spectral_analysis} we present our spectral analysis of nine BeXRB \nustar spectra and use our results to investigate a potential relation between the hard spectral component and the magnetic field strength in Sect.~\ref{sec:obs:magnetic_field}.

\subsection{Source selection}\label{sec:sourceselection}
We carried out a literature study and archive search for all BeXRBs to determine whether the sources are known to enter a state of stable quiescent accretion and, if so, whether archival broadband \nustar observations are available.
For many systems such as \tworxp, GRO~J2058+42, IGR~J18027-2016, RX~J0440.9+4431, and SAX J2103.5+4545, we find that the systems have been observed by \nustar only at luminosities exceeding $10^{35}\,\mathrm{erg}\,\mathrm{s}^{-1}$ \citep{krivonos2015,salganik2025,lutovinov2017,salganik2023,brumback2018}, and they are therefore not considered further in this work. 
The \nustar observations of \maxiOne and MXB~0656-072 indicate that the systems could exhibit X-ray emission caused by stable low-level accretion; the \nustar data are, however, insufficient to conduct a broadband spectral analysis with a focus on the intermediate and hard X-ray band \citep{ghimiray2024,raman2023}; they were therefore discarded from the sample.

In summary, we found nine BeXRBs (\twoSXPS\footnote{This object was initially detected and classified as a soft gamma repeater \citep{barthelmy2016} under the name \sgr.
Magnetars are known to exhibit bursts associated with crust fractures \citep{thompson1995}.
\citet{doroshenko2021} show concrete evidence for pulsations, which were also already hinted at by \citet{harrison2017}. The former argue that the X-ray source observed by \nustar is most likely not the counterpart of the burst and is actually a BeXRB, which they refer to as \twoSXPS.}\citep{doroshenko2021}, \aOF \citep{ballhausen2017,tsygankov2019a}, \cep \citep{vybornov2017,sokolova-lapa2023a}, \gro \citep{lutovinov2021,chen2021}, \gxTOF \citep{roucoescorial2018,tsygankov2019,sokolova-lapa2021}, \igr \citep{gorban2022a,ghising2023a}, \maxiZero \citep{pike2023,rai2023}, \srga \citep{doroshenko2022}, and \xper \citep{mushtukov2023,rodi2024,rai2025}),
with \nustar observations at low luminosities of $10^{33}$--$10^{34}\,\mathrm{erg}\,\mathrm{s}^{-1}$ that indicate stable quiescent accretion.
We based our subsequent spectral analysis on the \nustar observations of these nine BeXRBs.

\subsection{Semi-physical and phenomenological spectral models}\label{sec:obs:models}
The double humped X-ray spectra of BeXRBs at luminosities of ${\sim}10^{34}\,\mathrm{erg}\,\mathrm{s}^{-1}$ are often described as the sum of two semi-physical models for thermal Comptonization spectra such as  \texttt{comptt} \citep{titarchuk1994}. Examples of the application of such models are observations of \gro \citep{lutovinov2021}, \aOF \citep{tsygankov2019a}, \xper \citep{rai2025}, and \maxiZero \citep{malacaria2026}. While describing the spectrum of the low-energy component (LEC) with Comptonization is physically motivated, this is less clear for the case of the high-energy component (HEC), which is observed in an energy band where other effects than thermal Comptonization are expected to be at work.

Here, we model the X-ray spectra of BeXRBs in quiescence with a purely empirical model that focuses on measuring the position of the HEC without any physical interpretation.
Our primary aim is to describe the spectral positions of the two humps. Since we only want to constrain the continuum shape, we only considered the X-ray spectrum up to the red wing of the HEC and disregard the fundamental CRSF and its blue wing.
We modeled the LEC and the red wing of the HEC each with a power law with an exponential cut off (\texttt{cutoffpl} in \texttt{XSPEC} and \texttt{ISIS}).
We expected our data above ${\sim}30$\,keV to be poorly constrained for two reasons.
First, the \nustar effective area significantly decreases toward high energies \citep[see, e.g.,][Fig.~2]{harrison2013}.
Second, previous observations and their analyses as well as theoretical spectral models predict that the HEC is imprinted with a deep, wide cyclotron line \citep[see, e.g.,][]{tsygankov2019a,sokolova-lapa2021}.
While the line itself cannot be resolved in quiescent \nustar observations if $E_\mathrm{CRSF}\gtrsim 45$\,keV, the absorption and spectral redistribution caused by the cyclotron interaction result in a flux decrease toward higher energies.
As a result, typical \nustar spectra with exposure times of tens of~kiloseconds do not allow us to constrain the HEC reliably, which potentially results in model degeneracies between the photon index $\Gamma_2$ and the folding energy $E_\mathrm{fold}$, which determines the exponential rollover.
We therefore decided to fix the photon index of the HEC-modeling power law component $\Gamma_2$.

While models with  $\Gamma_2>0$ generally obtain favorable fit statistics, the spectral models lack physical interpretability due to the fact that in this case, the two \texttt{cutoffpl} components no longer model the two humps, respectively. 
Additionally, models where $\Gamma_2>0$ imply the unphysical scenario that the two components overlap significantly and that the HEC contributes a few percent of the flux in the soft band. 
This is not expected based on our theoretical understanding of the spectral formation within inhomogeneous neutron star atmospheres \citep[see, e.g.,][Fig.~4]{sokolova-lapa2021}

A preliminary analysis of \xper showed that fixing $\Gamma_2=-2$ is a balance between obtaining a statistically satisfactory description of the data and obtaining a description where each \texttt{cutoffpl} component models one spectral hump, respectively.
A more detailed justification is given in Sect.~\ref{sec:disc:tracer}.
From a physical point of view, the HEC-modeling component represents a Wien-like continuum in this case \citep[see, e.g.,][]{nagel1981,orlandini2001}.
In summary, the continuum model can be expressed as
\begin{equation}\label{eq:twohump}
	F_\mathrm{ph}(E)=\underbrace{K_1 \cdot E^{-\Gamma_1} \cdot
          \exp
          \left[-\frac{E}{E_\mathrm{fold,1}}\right]}_\text{LEC}+\underbrace{K_2\cdot
          E^2 \cdot
          \exp\left[-\frac{E}{E_\mathrm{fold,2}}\right]}_\text{Wien-like
          red wing of the HEC},
\end{equation}
where $K_{1,2}$ are the normalizations of the two components, $\Gamma_1$ and $E_\mathrm{fold,1}$ are the photon index and the folding energy of the LEC, and $E_\mathrm{fold,2}$ is the characteristic energy of the Wien-like component which determines the position and slope of its exponential cut off.

In order to characterize the relative positions of the two spectral components we reparametrize Eq.~\eqref{eq:twohump} by introducing their intersection energy, $E_\mathrm{int}$, defined as the energy at which the flux of the two \texttt{cutoffpl} components is equal, which replaces $K_2$ as a model parameter.
$K_2$ is therefore given by 
\begin{equation}
	K_2=K_1 \times E_\mathrm{int}^{-(\Gamma_1+2)} \times \exp\left[E_\mathrm{int} \times \left(\frac{1}{E_\mathrm{fold,2}}-\frac{1}{E_\mathrm{fold,1}}\right)\right].
	\label{eq:k2calc}
\end{equation}

\subsection{Spectral analysis utilizing the \doublehump model}\label{sec:spectral_analysis}
We based our spectral analysis on archival observations of nine BeXRBs at low luminosities observed with \nustar \citep{harrison2013}. This mission offers the broad band X-ray coverage needed and provides a sufficiently large number of archival observations. 
See Table~\ref{tab:app1} for the parameters of these observations. 
Where available, we also utilize contemporaneous observations or observations at a similar luminosity level with  \xrt \citep{gehrels2004,burrows2005}, \xmm \citep{jansen2001,denherder2001,struder2001,turner2001}, \nicer \citep{gendreau2012,gendreau2016}, and \chandra \citep{weisskopf2000}. 

\begin{figure*}
\centering
	\includegraphics[width=\textwidth]{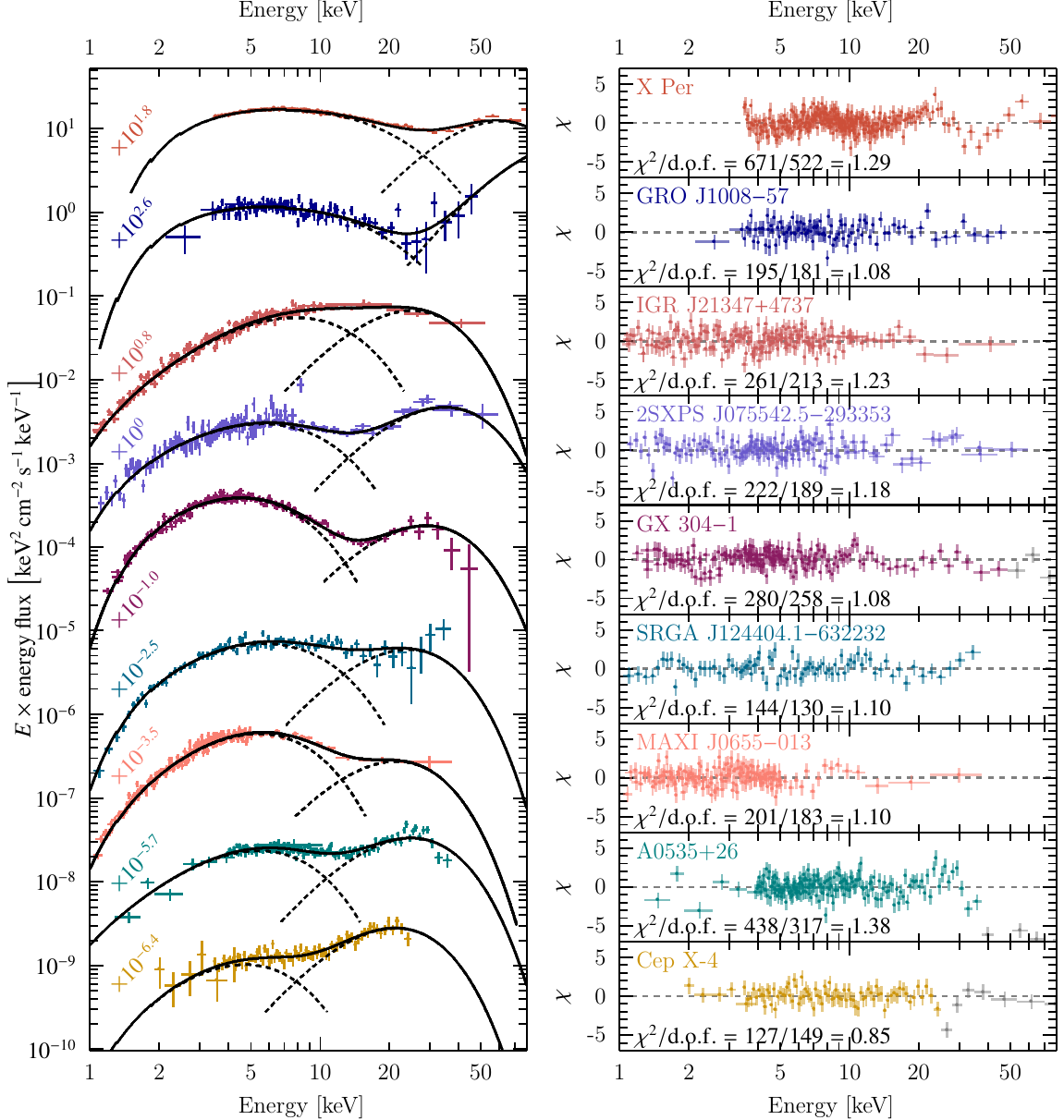}
	\caption{Spectral fit of the \doublehump model to quiescent X-ray spectra of nine BeXRBs. Left panel: Unfolded $E \times F_E$ spectra and fit model (solid line); the dashed lines indicate the two components of the \doublehump model. The spectra are separated vertically for improved perceptibility, ordered by descending intersection energy from top to bottom, and to this end have been rescaled by the factors given on the left $y$-axis of the left panel.  The spectra have also been rescaled to match \nustar/FPMA according to the  cross-calibration constants. Right panel: $\chi$ residuals of spectral fits for each source, respectively. The gray residuals indicate data points that were not used for spectral fitting, as they are at or above the CRSF feature, which is not part of the \doublehump model and therefore excluded. The spectra have been rebinned and the \nustar FPMs combined for visual purposes.}
	\label{fig:doublehump}
\end{figure*}

As the \doublehump model does not account for absorption features due to cyclotron resonant scattering, it naturally cannot reproduce CRSFs. Therefore, the X-ray spectra of sources exhibiting such features are only modeled up to energies below the CRSF, that is, to 40\,keV for \aOF, 50\,keV for \gxTOF, and 25\,keV for \cep.\footnote{For transparency, in Fig.~\ref{fig:doublehump} we nevertheless show the spectral fit for the full \nustar energy range, coloring data points that have not been included in the fit in gray.}
We utilized the absorption model \texttt{TBabs} \citep{wilms2000} with cross sections and abundances as given by \citet{verner1996} and \citet{wilms2000}, respectively.
A preliminary study showed that the absorption column density, $N_\mathrm{H}$, can be constrained for all BeXRBs considered,  with the exception of \cep, for which we assumed the Galactic \ion{H}{i} column density  in the direction of the source \citep{hi4picollaboration2016}.
Including a cross-calibration constant for each detector, fixed to unity for \nustar/FPMA, the model applied to each dataset is
\begin{equation}
	I(E)=\texttt{detconst} \times \texttt{TBabs}(E) \times \texttt{doublehump}(E).
	\label{eq:fullModel}
\end{equation}
Our spectral fits and their residuals are shown in Fig.~\ref{fig:doublehump}, the fit parameters are given in the lower half of Table~\ref{tab:app1}.

The \doublehump model describes the X-ray spectra of the nine selected BeXRBs comparably well.
For \gro, \igr, \twoSXPS, \gxTOF, \srga and \maxiZero the residuals are predominantly scattered around zero with some of the showing minor deviations toward higher energies, which is to be expected from both our stiff modeling of the HEC and the poor \nustar data constraints at high energies.
For \aOF and \cep dips in the residuals around 40\,keV and 25\,keV are apparent, which are due to their well-known cyclotron lines \citep[see, e.g.,][]{ballhausen2017,furst2018}, which are not modeled by the \doublehump model.
For \xper we see a wave-like structure at 30\,keV, which is potentially also related to stiff modeling  of the red wing  of the HEC, i.e., fixing $\Gamma_2=-2$.

While the \doublehump model is generally able to describe the X-ray spectra of the selected sources well, the question whether the sources are in the low-luminosity state as described in Sect.~\ref{sec:introduction} remains.
We assess this based on two criteria.
First, we inspect the $E\times F_E$ spectra visually and find that the spectra show strong indications of two individual components for all sources except for \igr, whose X-ray spectrum is noticeably flatter than of the other eight sources.
For the eight other BeXRBs, the dashed lines in Fig.~\ref{fig:doublehump} indicate that the two model components each describe the physical component they are intended to model and intersect at intermediate energies.

As a second test, we also fit a simple cut off power law to all sources and compare its $\chi^2$ fit statistic to that of the \doublehump fit; the improvement in fit statistic, $\Delta \chi^2$, is given in the lowest row of Table~\ref{tab:app1}.
For all nine BeXRBs this simple test provides a further argument that they exhibit hard X-ray emission beyond a simple cut off power law:
first, if the sources did not have statistically significant hard X-ray emission, the fit algorithm would push the normalization of the hard \texttt{cutoffpl} component, $K_2$, toward 0, which in our reparametrized form implies that $E_\mathrm{int}$ would be pushed toward high values.
Second, $\Delta \chi^2$ would approach 0.
As both are not the case, this simple comparison to a pure cut off power law fit shows that all nine sources exhibit hard X-ray emission in excess of a cut off power law.

The intersection energies of the two spectral components are between 7--28\,keV and are generally well constrained. 
$E_\mathrm{int}$ shows no correlation with X-ray luminosity.
We base our following subsequent analysis on these spectral fits, with a particular focus on the intersection energy.

\subsection{Relating the intersection energy to the magnetic field strength}\label{sec:obs:magnetic_field}
Having obtained a satisfactory description of a small sample of X-ray spectra of BeXRBs in quiescence, we now investigate the potential relation between magnetic field strength and intersection energy.
Secure CRSF detections in outburst exist for four out of the nine BeXRBs considered here, \aOF, \cep, \gro, and \gxTOF.
We estimate the cyclotron line energy based on data collected by \citet[][see their Fig.~9]{staubert2019}; for \gro we rely on data by \citet{chen2021}.
The data for \aOF stem from \citet{caballero2007} and  \citet{sartore2015}, for \cep from \citet{furst2015} and for \gxTOF from \citet{yamamoto2011} and \citet{malacaria2015}.
\begin{figure}
\centering
	\includegraphics[width=\columnwidth]{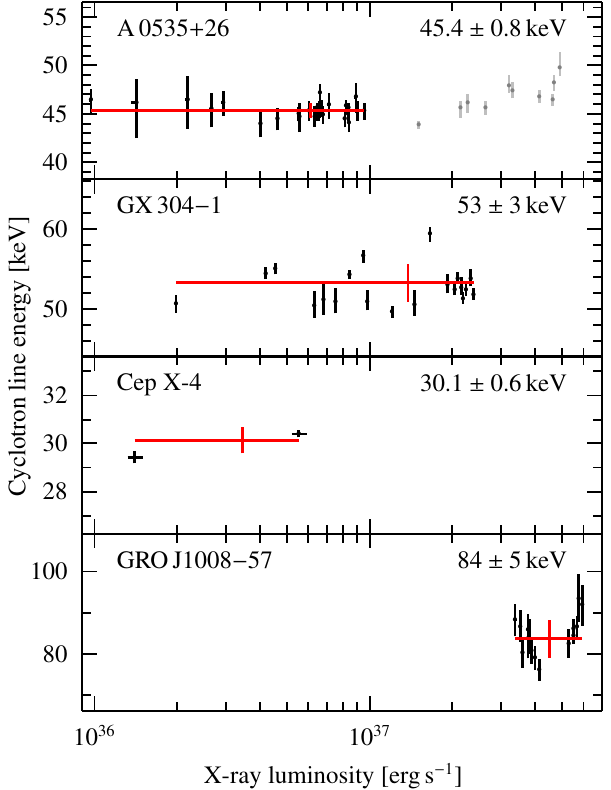}
	\caption{Observed cyclotron line energy as a function of X-ray luminosity. Red data points denote the inverse-variance weighted average of the cyclotron line energy. Gray data points have been excluded from the averaging. The data for the top three panels have been collected by \citet{staubert2019} and stem from \citet{caballero2007}, \citet{sartore2015} and \citet{furst2015}. The data for the bottom panel have been directly obtained from \citet{chen2021}.}
	\label{fig:ecyc_data}
\end{figure}

For each of the four sources, we selected cyclotron line energy values measured at X-ray luminosities between $10^{36}\,\mathrm{erg}\,\mathrm{s}^{-1}$ and a few $10^{37}\,\mathrm{erg}\,\mathrm{s}^{-1}$, i.e., for a luminosity range below the highest outburst luminosities, where radiation pressure effects can significantly affect the location of the region where the line is formed.
We computed their weighted average, using the inverse variance as weights, and determined the standard deviation, derived from their variance, and used them for our subsequent analysis.
The data and the obtained $E_\mathrm{CRSF}$ values are shown in Fig.~\ref{fig:ecyc_data}.

\begin{figure}
\centering
	\includegraphics[width=\columnwidth]{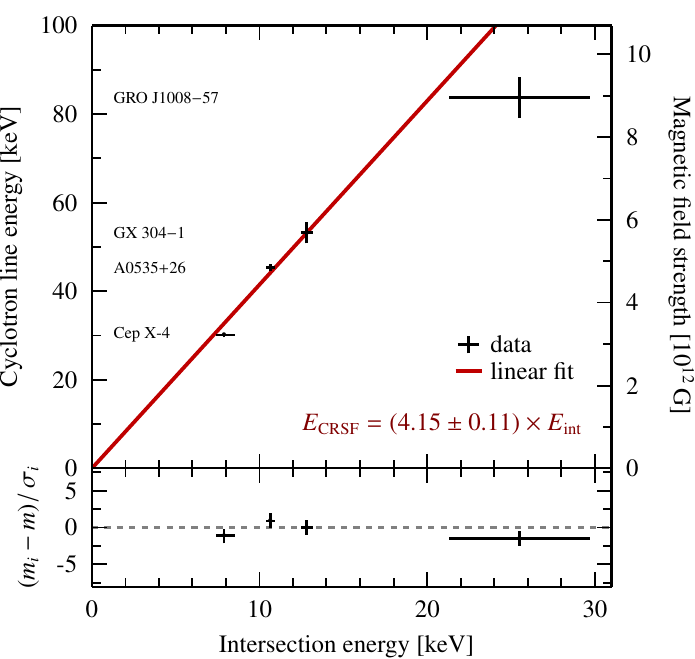}
	\caption{Relation between magnetic field strength and
          intersection energy. Top panel: cyclotron line energies from literature versus intersection energies as obtained in this work.
		  Black data points: sources with
          magnetic fields known from CRSF observations. Red line:
          linear regression taking uncertainties of $E_\mathrm{CRSF}$
          and $E_\mathrm{int}$ into account that is based
          on the four data points, with the relation is given in the
          lower right corner of the panel. Bottom panel: Residual of
          the best fit in units of the individual ratio uncertainty,
          $\sigma_i$.}
	\label{fig:ecrsf_intersect_DP}
\end{figure}
\begin{table}
	\renewcommand{\arraystretch}{1.2}

	\caption{Outburst cyclotron line energies, intersection energies, and their ratio for the four sources with confirmed CRSFs.}
	\begin{tabular}{lrrr}
		\hline \hline Source & $E_\mathrm{CRSF}$\,[keV] & $E_\mathrm{int}$\,[keV] & Ratio \\ \hline
		GX 304$-$1 & $53\pm3$ & $12.8\pm0.4$ & $4.1\pm0.3$\\ 
		A0535+26 & $45.4\pm0.8$ & $10.6\pm0.3$ & $4.3\pm0.2$\\ 
		GRO J1008$-$57 & $84\pm5$ & $25\pm5$ & $3.3\pm0.6$\\ 
		Cep X-4 & $30.1\pm0.6$ & $7.9^{+0.7}_{-0.5}$ & $3.8\pm0.3$ \\ \hline
	\end{tabular}
	\label{tab:ratios}
\end{table}
In Fig.~\ref{fig:ecrsf_intersect_DP}, we depict the variation of the cyclotron line energy versus the measured intersection energy, for the four sources. The data suggest a positive correlation between the two quantities.
The corresponding data is given in Table~\ref{tab:ratios}.
As both $E_\mathrm{int}$ and $E_\mathrm{CRSF}$ are subject to statistical and systematic uncertainties, one cannot model their relation simply by fitting a model and minimizing the $\chi^2$~statistic as it does not take the uncertainty of the independent coordinate into account. 
Instead, we calculated the ratio between outburst cyclotron line energy and intersection energy for each of the four data points.
The uncertainty is estimated via Gaussian error propagation.
From the four ratios, we calculate the weighted average with the inverse variance of the ratios as weights.
We obtained 
\begin{equation}
	E_\mathrm{CRSF} = (4.15\pm0.11) \times E_\mathrm{int}.
	\label{eq:ecrsf_relation}
\end{equation}
One can make use of the relation between cyclotron line and magnetic field strength \citep[see, e.g.,][Eq.~1]{staubert2019} to express the magnetic field strength as a function of intersection energy as
\begin{equation}
	B/10^{12}\,\mathrm{G} = (1+z) \times (0.358\pm0.009) \times E_\mathrm{int}/\mathrm{keV}.
\end{equation}
Further assuming a redshift of $z=0.24$, which is expected for a $1.4\,M_\odot$ neutron star with a radius of 12\,km, we obtained
\begin{equation}
	B/10^{12}\,\mathrm{G} = (0.444\pm0.012) \times E_\mathrm{int}/\mathrm{keV}.
	\label{eq:bmagz124_relation}
\end{equation}
For a canonical neutron star in a BeXRB system, the intersection energy in keV can therefore be roughly estimated as twice the magnetic field strength in units of $10^{12}$\,G, which we designate as the 2-$B$-12 rule, analogous to the well-known 12-$B$-12 rule.
In summary, we have shown that there exists a positive correlation between the surface magnetic field strength of neutron stars in BeXRBs and the intersection energy between the two spectral components present in their low-luminosity X-ray spectra.

\section{Application of the \doublehump model to simulated BeXRB spectra}\label{sec:theory}
The results presented in Sect.~\ref{sec:observation} indicate a positive correlation between magnetic field strength and intersection energy.
In this section we show that such a correlation is also expected from a theoretical point of view. Specifically,  from the growing number of physical spectral models for the quiescent state of BeXRBs \citep[see, e.g.,][]{mushtukov2021} we make use of the polar cap model (\polcap hereafter) developed by \citet{sokolova-lapa2021}. We introduce this model and its parameterization in Sect.~\ref{sec:polcap} and show that the model indicates a relation between the hump position and the magnetic field. In Sect.~\ref{sec:simulation} we then simulate \nustar observations based on this model and show that the relation between magnetic field strength and intersection energy is in qualitative agreement with our \nustar observations.

\subsection{The \polcap model}\label{sec:polcap}
\begin{figure}
\centering
	\includegraphics[width=\columnwidth]{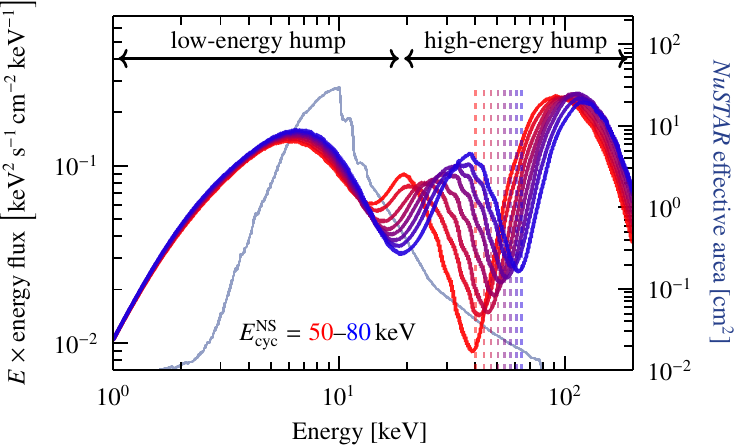}
	\caption{Spectral model \polcap for different magnetic field strengths. The magnetic field strengths are varied for cyclotron energies between 50--80\,keV in the NS frame. The X-ray spectra are shown in the observer's frame. The colored dashed lines indicate the cyclotron energy in the observer's frame. The green line shows the \nustar effective area for a source region with $30''$ diameter and $1'$ off-axis angle.}
	\label{fig:polcapVariations}
\end{figure}

The \polcap model \citep{sokolova-lapa2021} is a theoretical spectral model that aims to reproduce the X-ray spectra of highly magnetized neutron stars at luminosities of \mbox{$10^{33}$--$10^{34}\,\mathrm{erg}\,\mathrm{s}^{-1}$}.
It is based on a numerical solution of the polarized radiative transfer equation, from which the resulting emergent spectrum is derived for a variety of magnetic field strengths and mass accretion rate configurations.
The emergent spectra are tabulated for a range of configurations.
In its current realization, the \polcap model takes three input parameters:
\begin{enumerate}
	\item the model normalization $N$, proportional to the ratio between the full area, onto which matter is being accreted that covers all polar caps, $A_\mathrm{accretion}$, and squared distance of the observer, $D$,
		\begin{equation}
			N=\frac{A_\mathrm{accretion}/ 1\,\mathrm{m}^2}{\left(D/1\,\mathrm{kpc}\right)^2};
		\end{equation}
	\item the logarithm of the area density of the mass flux accreted onto a \textsl{single} pole 
		\begin{equation}
			\log \mathcal{F}_\mathrm{mass}\equiv\log \left[\left(\frac{\dot M}{1\,\mathrm{g}\,\mathrm{s}^{-1}}\right)\cdot \left(\frac{A_\mathrm{cap}}{1\,\mathrm{cm}^2}\right)^{-1}\right],
		\end{equation}
which is the logarithm of the ratio between mass accretion rate and area of \textsl{one} polar cap, onto which the matter is accreted; and 
	\item the cyclotron energy, $E^\mathrm{NS}_\mathrm{cyc}$, that is the energy where the cyclotron interaction cross section between photons and atmospheric electrons has its maximal value in the NS rest frame. The energy is naturally closely related to the cyclotron \textsl{line} energy of the observed absorption feature, which can be approximated by dividing the cyclotron energy by $(1+z)$ to compensate for the gravitational redshift. Where required, we convert the cyclotron energy to to the cyclotron \textsl{line} energy in the observer's frame assuming a gravitational redshift of $z=0.24$.
\end{enumerate}
In order to demonstrate how  the emergent spectrum changes with magnetic field strength, in Fig.~\ref{fig:polcapVariations} we show spectra computed with the \polcap model for various magnetic field strengths (identified via the cyclotron energy in the NS frame).
It is evident that the high-energy hump is shifted toward higher energies with increasing $B$-field strength, whereas the soft X-ray component shows a much weaker dependence on the magnetic field strength, which already suggests a positive correlation between the relative position of the two components and the magnetic field strength.
Figure~\ref{fig:polcapVariations} also shows the \nustar effective area, which suggests that the HEC and the CRSF imprinted on it most likely cannot be resolved in typical cases where $B \gtrsim \mathrm{a\:few}\,10^{12}$\,G due to the rapidly decreasing effective area toward high energies.

\subsection{Simulations of X-ray spectra of BeXRBs at low luminosities}\label{sec:simulation}
To study how the spectral changes with magnetic field predicted by the \polcap models are reflected in observations, we simulated 100\,ks \nustar observations of the spectra appropriate for BeXRBs located at a distance of 2\,kpc and at an observed luminosity of $10^{34}\,\mathrm{erg}\,\mathrm{s}^{-1}$. The instrument response files and background spectra are taken from the 2022 \nustar observation of \gxTOF (OBS\ ID 30701015002). 
The parameters of the \polcap model are chosen as follows.
The normalization is set such that the luminosity in the observer's frame is $10^{34}\,\mathrm{erg}\,\mathrm{s}^{-1}$.
We vary the mass flux over the whole tabulated range, that covers the whole range of values expected to prevail at low luminosities.
The cyclotron energies are 50, 60, 70 and 80\,keV in the NS rest frame, respectively, which covers the whole tabulated range. 

\begin{figure}
\centering
	\includegraphics[width=\columnwidth]{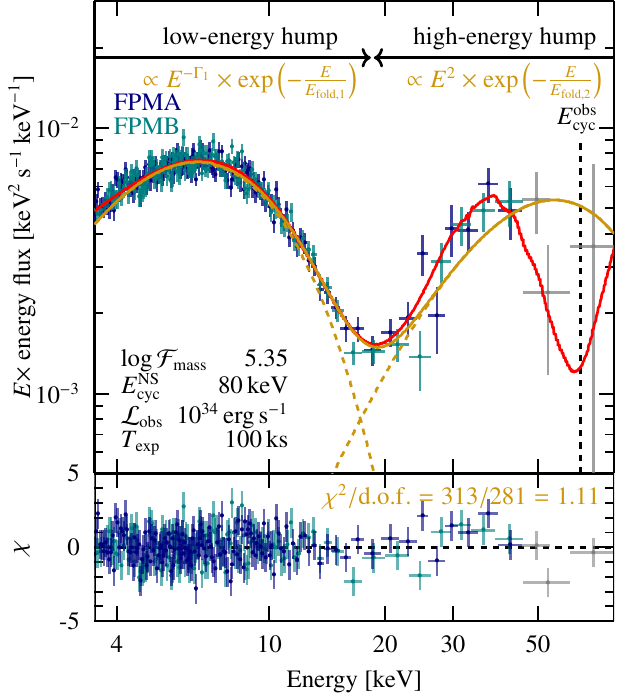}
	\caption{Simulated \nustar spectrum based on the \polcap model. Top panel: Simulated rebinned spectra in blue and teal for the two \nustar focal plane modules, respectively. The red line shows the \polcap model, on which the simulated spectra are based. The input parameters are given in the lower left corner. Consistent with our treatment in Sect.~\ref{sec:observation}, we base spectral fits only on data points sufficiently well below the CRSF energy, as the \doublehump model does not account for CRSFs; data points not used spectral fitting are shown in gray. The fit \doublehump model is shown in orange; its two \texttt{cutoffpl} components are shown as dashed orange lines. The cyclotron energy in observer's rest frame is indicated by the black dashed line. Bottom panel: $\chi$~residuals of \doublehump fit.}
	\label{fig:example}
\end{figure}

We then fit the simulated spectra, rebinned to 30\,counts per bin, with the \doublehump model (see Fig.~\ref{fig:example} for an example). 
As the \doublehump model only describes the LEC and the red wing of the HEC, only data points up to~$0.9\times E^\mathrm{obs}_\mathrm{cyc}$ are used to fit the model. The intersection energies obtained from the simulated spectra are shown as the dark green area in Fig.~\ref{fig:ecrsf_intersect_DPT}.
To extrapolate toward higher magnetic field strengths than those tabulated for the \polcap model ($E_\mathrm{cyc}^\mathrm{NS}=80\,\mathrm{keV}$, corresponding to ${\sim} 7 \times 10^{12}$\,G), the left and right boundaries of the parameter space, that is the dark green area in Fig.~\ref{fig:ecrsf_intersect_DPT}, were fit with straight lines through the origin, whose slopes are $m_\mathrm{left}=4.08$ and $m_\mathrm{right}=3.52$.

\subsection{Comparison between observations and theory}
\begin{figure}
\centering
	\includegraphics[width=\columnwidth]{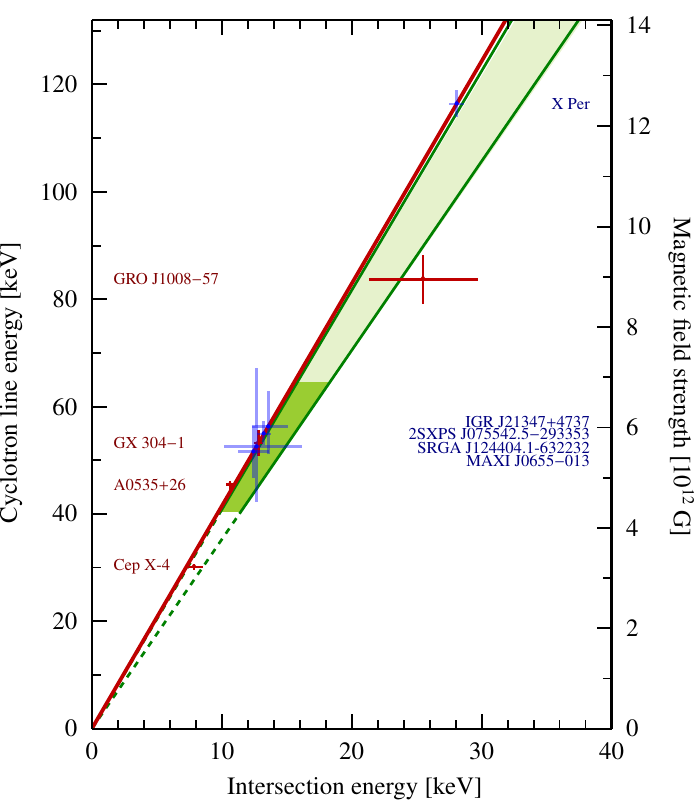}
	\caption{Relation between magnetic field strength and
          intersection energy. Shown in red are the data points from
          real observations with the corresponding source names given
          on the left $y$-axis. The red line indicates the linear relation found in Sect.~\ref{sec:obs:magnetic_field}. The blue data points indicate the \textsl{predicted} magnetic field strengths for those BeXRBs without known magnetic field strength. Their names are given on the right $y$-axis. The solid green lines are obtained from the simulation-driven approach and resemble the left and right boundaries of the parameter space; the dashed green line show the behavior toward low $B$-field strengths, for which the  \polcap is not applicable due to spectral redistribution due to Landau excitations. The dark green area indicates the $B$-field range from which the green lines are obtained from; the light green area shows the extrapolation of the dark green area toward higher $B$-field strengths.}
	\label{fig:ecrsf_intersect_DPT}
\end{figure}

In the sections above we obtained relations between the relative position of the two spectral components, quantified as $E_\mathrm{int}$, and the magnetic field strength in form of the cyclotron line energy from real observations and simulated spectra based on theoretical works, respectively.
To compare the relations, we show the linear fit to the data points from real observations from Fig.~\ref{fig:ecrsf_intersect_DP} and the parameter space obtained from simulated spectra together in Fig.~\ref{fig:ecrsf_intersect_DPT}.
The observation-driven relation from Sect.~\ref{sec:obs:magnetic_field} and the theory-driven results both indicate a positive correlation between magnetic field strength and intersection energy.
The data-driven relation is in good agreement with the left boundary of the parameter space given by the theory-driven approach.

Based on the data-driven linear correlation in Eqs.~\ref{eq:ecrsf_relation}--\ref{eq:bmagz124_relation}, we can also estimate the magnetic field strengths of the five BeXRBs studied in Sect.~\ref{sec:observation} for which no CRSFs have been claimed with high statistical significance, namely \twoSXPS, \igr, \maxiZero, \srga and \xper.
The predicted values are shown in Fig.~\ref{fig:ecrsf_intersect_DPT} as blue data points and are additionally listed in Table~\ref{tab:predicted}.
\begin{table}
	\setlength{\tabcolsep}{2.5pt}
	\caption{Predicted magnetic field strengths and cyclotron line energies based on the observation-driven linear relation between intersection energy and cyclotron line energy and assuming $z=0.24$.}
	\renewcommand{\arraystretch}{1.25}
	\begin{tabular}{lccc}
		\hline \hline 
		Source & $E_\mathrm{int}$ & $E_\mathrm{CRSF}^\mathrm{predicted}$ & $B^\mathrm{predicted}$ \\
		 & $\left[ \mathrm{keV} \right]$ & $\left[ \mathrm{keV} \right]$ & $\left[ 10^{12}\,\mathrm{G} \right]$ \\ \hline
		X Per & $28.1^{+0.7}_{-0.6}$ & $116\pm3$ & $12.5\pm0.3$ \\
		SRGA J124404.1-632232 & $13^{+4}_{-3}$ & $50\pm20$ & $6\pm2$ \\
		MAXI J0655$-$013 & $12\pm2$ & $52\pm5$ & $5.5\pm0.6$ \\
		IGR J21347$+$4737 & $14\pm2$ & $56^{+7}_{-6}$ & $6.0^{+0.7}_{-0.6}$ \\
		2SXPS J075542.5$-$293353 & $13.2\pm0.6$ & $55\pm3$ & $5.9\pm0.3$ \\ \hline
		\end{tabular}	
	\label{tab:predicted}
\end{table}
While these values should be understood only as rough estimates of the magnetic field strength, based on a linear regression to only four data points with prior assumptions (see the discussion of applicability and limitations of the \doublehump model in Sect.~\ref{sec:disc:applicability}) they serve as testable hypotheses that are verifiable or falsifiable by deep observations of the hard X-ray band that could provide the statistics to find high-energy cyclotron lines.

\section{Discussion}\label{sec:discussion}
In this work, we have shown that the intersection energy between the two humps present in X-ray spectra of BeXRBs with luminosities of ${\sim} 10^{34}\,\mathrm{erg}\,\mathrm{s}^{-1}$ is correlated with the surface magnetic field strength inferred from CRSF observations. In this section we discuss the consequence of this result. We start in Sect.~\ref{sec:disc:physics} with a discussion of the physical context. In Sect.~\ref{sec:disc:applicability}, we discuss the applicability and limitations of the \doublehump model and the linear correlation found in Sects.~\ref{sec:observation} and~\ref{sec:theory}.
In Sect.~\ref{sec:disc:tracer} we discuss and other possible tracers of the magnetic field strength in light of our findings.
Given its pivotal role in recent history and its highly interesting spectrum, we discuss \xper individually in Sect.~\ref{sec:disc:xper}.
Lastly, we outline possible next steps and give an outlook in Sect.~\ref{sec:disc:outlook}.

\subsection{Physical interpretation}\label{sec:disc:physics}
Our study of real and simulated BeXRB observations at low luminosity has shown a positive correlation between the relative position of the two spectral components present in their X-ray spectra and the NS magnetic field strength. As theoretical models generally assume that the LEC depends much less on the magnetic field than the HEC (see Fig.~\ref{fig:polcapVariations} for the example of the \polcap model), in the following we concentrate on the formation of the HEC. 

In the model of \citet{mushtukov2021}, the HEC forms in an overheated atmosphere with an effective conversion of the kinetic energy of infalling matter to cyclotron photons.
In this picture, the hump therefore consists of Comptonized cyclotron photons and is imprinted with a cyclotron (absorption) line that is due to inelastic scattering of cyclotron photons and electrons within the NS atmosphere.
In this picture, the energy of the reprocessed cyclotron photons, and therefore the general position of the HEC, depends directly on the magnetic field strength.

Alternatively, \citet{sokolova-lapa2021} suggest that the HEC forms within the NS atmosphere, where initially ambient polarized soft X-ray photons are reprocessed in Compton interactions with infalling particles. These interactions increase the photon energy to tens of~kiloelectronvolts.
Cyclotron interactions that effectively remove photons with $E_\gamma \sim E_\mathrm{cyc}$ alter the emergent spectral shape and imprint a deep absorption feature onto HEC.
The cyclotron line itself is naturally related to the magnetic field strength.
Therefore, the red wing of the CRSF, and, -- although not observable -- also its blue wing, are expected to depend on the magnetic field strength.
In our empirical modeling attempt, we focus on the red wing and especially its rising flank to identify its intersection energy, which is subsequently also expected to depend on the magnetic field strength.

While these two theoretical frameworks utilize slightly different mechanisms for the formation of the HEC, both of them predict that the HEC location, and therefore also the intersection energy when the humps are modeled with the \doublehump model, depends on the magnetic field. We directly demonstrated this for the physical model of \citet{sokolova-lapa2021} in Sect.~\ref{sec:theory}.

It is noteworthy that already much earlier theoretical studies suggested that signatures of the strong magnetic field of ${\sim} 10^{12}\,\mathrm{G}$ are present in the X-ray spectrum of HMXBs.
\citet{nelson1993} show that at least 10\% of the accretion energy should be converted to radiation in the hard X-ray band at energies \textsl{below} the cyclotron line energy.
They attribute this nonthermal emission to cyclotron photons, that are emitted in large numbers via the decay of electrons on higher Landau levels.
In inelastic scattering, the cyclotron photons on average decrease their energy, which results in hard X-ray emission at energies below the cyclotron line energy.
In a follow-up publication, \citet{nelson1995} find that between 0.5\% and 5\% of the accretion luminosity is expected to be emitted as cyclotron emission between 5--20\,keV. 
It is exactly this energy band that is crucial to determine the characteristics of the red wing of the HEC and the intersection energy of the two components.
Depending on the assumed physical processes that lead to the formation of the HEC, this emission at intermediate energies could indeed be understood as the result of cyclotron emission.

\subsection{The applicability of the \doublehump model and our relation derived from it}\label{sec:disc:applicability}
The \doublehump model itself is applicable to the X-ray spectra of BeXRBs with mass accretion rates low enough so that the accreted matter is not decelerated in a radiative shock, but within the NS atmosphere, such that double-humped X-ray spectra are formed.
Unfortunately, it is not possible to give a simple luminosity threshold for its applicability. Instead, one needs to verify the presence of a hard X-ray component for example by applying a simpler cut off power law model; if a statistically significant excess at intermediate to high energies remains, this is an indicator that the spectrum potentially exhibits two distinct spectral components, in which case the \doublehump model is applicable.    

We emphasize that systems harboring rapidly rotating neutron stars, with periods on the order of a few seconds, are not expected to enter a state of low-level, stable accretion at luminosities of ${\sim}10^{34}\,\mathrm{erg}\,\mathrm{s}^{-1}$.
Rather, the rotating magnetic field is expected to inhibit accretion onto the NS \citep{campana2002}.
As a result, these objects are not expected to show a double-humped X-ray spectrum and are thought to enter a quiescent state \citep[see, e.g.,][]{roucoescorial2020}.
Therefore, the \doublehump model is not applicable to them.

The applicability of the \doublehump model also depends on the magnetic field strength of the BeXRB in question.
Due to the influence of Landau excitations which alter the spectral shape to a rather flat shape without two humps \citep[see][Fig.~5a]{gorban2022} it is not valid for BeXRBs with field strengths below $2.5\times 10^{12}$\,G. 
In addition, as the \doublehump model only describes the red wing of the HEC, it cannot model cyclotron lines and can therefore only be utilized for X-ray spectra up to the energies that are not affected by cyclotron interactions. 
As the cyclotron line begins to significantly alter the continuum for $E \gtrsim 0.9 \times E_\mathrm{CRSF}^\mathrm{obs}$ \citep[see, e.g.,][]{meszaros1985a,sokolova-lapa2021}, we use this threshold as the upper energy limit for modeling using the \doublehump model.

Lastly, the spectral range of the data and the data's level of constraint also impact the parameters of the \doublehump model.
For poor data the intersection energy is naturally less well constrained and is potentially correlated with the other \doublehump parameters.
For example, a preliminary study of \maxiZero has shown that the intersection energy decreases by 5\,keV when utilizing well-constrained \xmm data instead of poorly constrained \xrt data along with 
\nustar data.
While the value found using \xmm is still within uncertainty of the \xrt-derived value, this example shows that poor soft X-ray data can alter the modeling of the soft X-ray component and subsequently the intersection energy.
It is therefore advisable to utilize \nustar in combination with highly sensitive soft X-ray instruments such as \chandra, \nicer or \xmm to achieve the most accurate spectral description possible. 

\subsection{Connecting the \doublehump model and the magnetic field}\label{sec:disc:tracer}
In our empirical \doublehump model, we have chosen to quantify the relative position of the two components via the intersection energy of the two cut off power laws utilized to describe the low-energy hump and the red wing of the high-energy hump.
As described in Sect.~\ref{sec:obs:models}, the reparametrization of the two \texttt{cutoffpl} components ensures that the components \textsl{are forced to} have an intersection point.
This approach already constitutes a restriction of the spectral model that potentially eliminates a statistically more favorable description of the data, where the two components do not intersect.

\begin{figure}
\centering
	\includegraphics[width=\columnwidth]{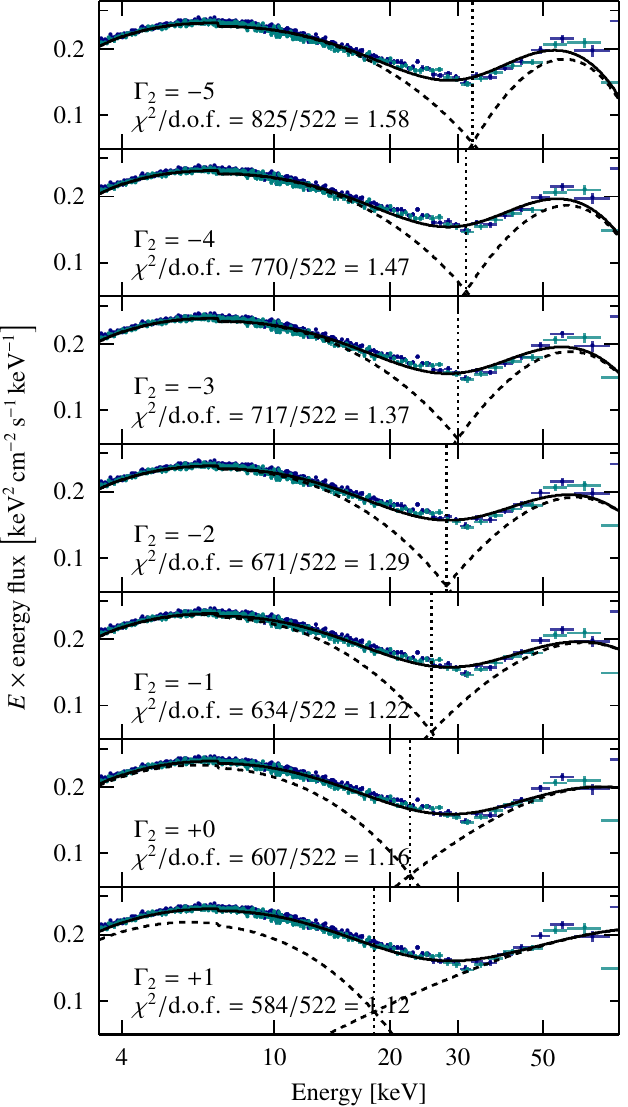}
	\caption{Impact of $\Gamma_2$ on spectral modeling. Each panel shows a fit to the X-ray spectrum of \xper (shown in blue and teal) with $\Gamma_2$ fixed to the values given in the panels. The solid line shows the best-fit model, the dashed lines indicate the two model components, the vertical dotted lines indicate the resulting intersection energy. The $\chi^2$~statistic is given in each panel.}
	\label{fig:xper_beta_panel}
\end{figure}

The restriction to $\Gamma_2=-2$ for the component describing the HEC poses a further restriction for the model that is not necessarily required for those sources with high enough signal-to-noise ratio to constrain the falling flank toward the cyclotron line of the red wing sufficiently well.
This is supported by directly fitting the \polcap model spectra, where the HEC can be better constrained as its importance is not reduced by  the rapidly declining effective area toward high energies present in real instruments. 
In this case we can constrain the full red wing of the HEC and generally find $\Gamma_2<0$, which
 is not feasible for real or simulated \nustar spectra.
As already touched upon in Sect.~\ref{sec:obs:models}, the restriction also serves the purpose to achieve a description where each cut off power law component models each hump, respectively, instead of modeling both humps at the same time. 
To illustrate this, we show fits to the X-ray spectrum of \xper for a variety of fixed values of $\Gamma_2$ in Fig.~\ref{fig:xper_beta_panel}.
Fits with negative $\Gamma_2$ achieve the desired model description, where each cut off power law component models one spectral hump, whereas those with positive $\Gamma_2$ achieve a favorable description only in a statistical sense, which nevertheless lacks physical interpretability.
We therefore conducted our spectral analysis in this work with a fixed value of $\Gamma_2=-2$.

While the intersection energy is arguably the most accessible parameter connecting the spectral shape and the magnetic field, as it only requires constraining the LEC and the rising flank of the HEC, other spectral parameters could in theory also be utilized.
For example, the energy where the red wing of the HEC reaches its maximum value or the ratio between the maxima of the HEC and LEC should also be considered as tracers of the magnetic field.
Both the position of the peak and the maximum flux value of the red wing of the HEC are inaccessible to \nustar at the typical magnetic field strengths considered here, $B\gtrsim \mathrm{a\:few}\,10^{12}$\,G.
Moreover, the theoretical reasoning why the ratio between the maxima should be related to the magnetic field strength is challenging.

Other spectral parameters have already been shown to correlate with the NS magnetic field strength  for HMXBs at much higher luminosities: \citet{makishima1999} and \citet{coburn2002} show that the cyclotron line energy is positively correlated with the spectral cut off energy. 
This correlation has also been found by the more recent study by \citet{pradhan2021}.
\citeauthor{coburn2002} propose that this could indeed be a physical correlation between the $B$-field strength and the cut off energy.
Alternatively, their positive correlation could also stem from a dependence of the cut off energy on the electron temperature, with the latter being potentially related to the magnetic field strength. 
The connection between $E_\mathrm{int}$ and magnetic field strength from this work on the other hand has a much more direct physical connection: the $B$-field dependence of the position of the HEC directly results in a measurable $B$-field dependence of the intersection energy.

\subsection{The case of \xper}\label{sec:disc:xper}
\xper is one of the most interesting sources touched upon in this work in many ways.
First, it is the only \textsl{persistent} BeXRB in this work, whereas the other sources are \textsl{transient} sources, that are mostly only detectable during outbursts by monitors. 
Second, it is by far the closest source with a photo geometric distance of $d\approx 0.6$\,kpc \citep{bailer-jones2021,gaiacollaboration2016,gaiacollaboration2023}.
It is therefore also the source with the highest flux in our sample and its \nustar observation subsequently provides the highest signal-to-noise ratio.
Third, \xper showcases how the scientific view of the quiescent state of BeXRBs has changed over the past 50\,years.
\xper was discovered by \citet{giacconi1972} with the \textsl{Uhuru} satellite; its optical counterpart was identified as a Be~star shortly thereafter \citep{moffat1973}.
\citet{white1976a} first detected X-ray pulsations with a periodicity of 836\,s.
Around 20~years later, \citet{disalvo1998} report a 0.1--100\,keV luminosity of ${\sim}2.4\times10^{34}\,\mathrm{erg}\,\mathrm{s}^{-1}$ based on a deep \textsl{BeppoSAX} observation.
While they find that the soft X-ray spectrum is well-described by a cut off power law, a hard excess remains.
Notably similar to the model utilized here, \citeauthor{disalvo1998} describe the hard excess with a second cut off power law and theorize, motivated by \citet{nelson1993}, and \citet{nelson1995}, that the hard excess could be related to cyclotron emission.
They approximate the cyclotron energy as the cut off energy of ${\sim}$65\,keV, which yields a $B$-field strength of ${\sim} 5.6\times 10^{12}$\,G (see last paragraph of Sect.~\ref{sec:disc:tracer}).

Shortly thereafter, \citet{coburn2001} employ a model consisting of a blackbody and cut off power law component and report that a CRSF situated at ${\sim}29$\,keV is required to obtain a satisfactory description of the RXTE data.
This potential cyclotron line implies a magnetic field strength of ${\sim}3\times10^{12}$\,G.
While the residuals of fits without the CRSF show a wave-like structure, this is only partly resolved by inclusion of the CRSF, which reduces the amplitude of the wave-like structure, but does not result in residuals statistically scattered around the zero-line \citep[see][Fig.~4]{coburn2001}.

The claim of a ${\sim}$30\,keV CRSF was reiterated by multiple later publications \citep[e.g.,][]{lutovinov2012,rodi2024,rai2025}; however the evidence that the apparent absorption feature is in fact not caused by cyclotron resonant scattering has been growing for the past 15 years, starting with \citet{doroshenko2012}, who modeled the \integral spectrum of \xper with two Comptonizing plasma components \citep[see][]{titarchuk1994} and interpreted the observed dip at ${\sim}30$\,keV as the dip between the two components and not as a cyclotron line.
Based on the earlier timing analysis by \citet{delgado-marti2001} and the spin-down mechanism proposed by \citet{illarionov1990}, they estimated the magnetic field of \xper to be  ${\sim}10^{14}$\,G, putting it within the range of magnetars \citep[see, e.g.,][]{olausen2014}.
The application of the \citet{ghosh1979} torque model by \citet{yatabe2018} also yielded a value in excess of ${\sim}10^{13}$\,G. These measurements are further arguments against interpreting the feature at 30\,keV as a CRSF, which would imply a field of  $B\sim3\times10^{12}$\,G. 

Moreover, the interpretation of the trough as the dip between the two spectral components has gained a much stronger theoretical foundation \citep{mushtukov2021,sokolova-lapa2021}, as such dips are expected at energies of a few tens of keV.
Recent \textsl{IXPE} observations of \xper reveal that it may be an orthogonal rotator and the results of \citet{mushtukov2023} are generally in agreement with the theoretical studies of polarized emission at low luminosities \citep[see, e.g.,][]{mushtukov2021,sokolova-lapa2021,sokolova-lapa2023b}.
While the direct confirmation of \xper's strong magnetic field of ${\gtrsim}10^{13}$\,G via CRSF detection above 100\,keV is not feasible with the current generation of X-ray instruments, future missions with increased sensitivity in the hard X-ray band and soft Gamma-ray band may be able to achieve this goal.
In this sense, \xper serves as an example that not all dips in the spectra of accreting neutron stars should automatically be interpreted as cyclotron lines.

\subsection{Next steps and outlook}\label{sec:disc:outlook}
We have shown that the NS surface magnetic field strength is positively correlated with the intersection energy between the two spectral components of BeXRBs at low luminosities of ${\sim}10^{34}\,\mathrm{erg}\,\mathrm{s}^{-1}$.
Based on the availability of observations of sources with known $B$-field strength and the $B$-field range for which the \polcap model has been tabulated for, we have investigated the relation for magnetic field strengths between a few~$10^{12}$\,G to approximately $10^{13}$\,G, roughly covering half an order of magnitude.

To determine through observations whether this relation can be extended remains an interesting open question. This question can only be answered by obtaining BeXRB spectra at low luminosities with independently verified magnetic field strengths.
Given the small number of accreting neutron stars in BeXRBs with known magnetic fields \citep[see][Tables A1 and A2]{staubert2019}, extending our sample further is challenging. 
Observations during outbursts pose the potential to measure the magnetic field strength via CRSF identification, for example; observations during the decline of the outburst potentially allow study of the system at low mass accretion rates associated with the double-humped spectral shape.
Each such pairing of measured magnetic field strength and observation during the low-luminosity state will help refine and possibly extend the validity of the linear relation probed here. All-sky surveys such as those of \textit{eROSITA} \citep[see, e.g.,][]{predehl2021,merloni2024} also have the potential to detect BeXRBs at  low luminosities due to ever-increasing survey sensitivity.

Another interesting aspect to pursue is the pulse-phase resolved analysis at low luminosities, which could show whether the HEC changes as a function of pulse phase due to the varying viewing angle of the anisotropically emitting emission region.
Analogous to the pulse-phase variability of CRSFs \citep[see, e.g.,][]{kreykenbohm2004,staubert2019,zalot2024}, an investigation of the variability of the intersection energy could strengthen or weaken our results. However, exposure times well in excess of 100\,ks would be required to study the phase-resolved spectra.

In parallel to these observational efforts, extensions of existing models  are required to understand whether the relative position of the two components and the magnetic field strength are in fact linearly related for a larger range of $B$-field strengths.

\begin{acknowledgements}
We acknowledge partial funding from Deutsche Forschungsgemeinschaft under contract 414059771. This research was done under the auspices of DFG Forschungsgruppe 2290 (eRO-STEP).
We acknowledge the pivotal contribution of Katja~Pottschmidt to this project, the research of highly magnetized accreting neutron stars, and especially her mentoring of so many students and early-career scientists on both sides of the Atlantic.
We deeply regret her untimely passing.
We are committed to upholding her lasting legacy in new scientific frontiers.
The authors thank Victoria Kaspi for her insight into \sgr and her helpful comments, Pragati Pradhan for her helpful comments, Rick Rothschild for his helpful comments on luminosity correlations, R\"udiger Staubert for providing us with cyclotron line energies that have been crucial for this work, Mihoko Yukita and the whole \nustar help desk for providing us with the names of observation PIs and the XMAG collaboration for productive discussions of this work.
This research has made use of a collection of ISIS functions (ISISscripts) provided by ECAP/Remeis observatory and MIT (\url{https://www.sternwarte.uni-erlangen.de/isis/}), data and software provided by the High Energy Astrophysics Science Archive Research Center (HEASARC), which is a service of the Astrophysics Science Division at NASA/GSFC and data from the European Space Agency (ESA) mission {\it Gaia} (\url{https://www.cosmos.esa.int/gaia}), processed by the {\it Gaia} Data Processing and Analysis Consortium (DPAC, \url{https://www.cosmos.esa.int/web/gaia/dpac/consortium}).
Funding for the DPAC has been provided by national institutions, in particular the institutions participating in the {\it Gaia} Multilateral Agreement.
\end{acknowledgements}

\bibliographystyle{aa}
\bibliography{references.bib}

@article{araya1999,
  title = {Cyclotron Line Features from Near-Critical Magnetic Fields: {{The}} Effect of Optical Depth and Plasma Geometry},
  author = {Araya, R. A. and Harding, A. K.},
  year = {1999},
  month = may,
  journal = {ApJ},
  volume = {517},
  pages = {334--354},
  doi = {10.1086/307157}
}

@article{bailer-jones2021,
  title = {Estimating {{Distances}} from {{Parallaxes}}. {{V}}. {{Geometric}} and {{Photogeometric Distances}} to 1.47 {{Billion Stars}} in {{Gaia Early Data Release}} 3},
  author = {{Bailer-Jones}, C. A. L. and Rybizki, J. and Fouesneau, M. and Demleitner, M. and Andrae, R.},
  year = {2021},
  month = mar,
  journal = {AJ},
  volume = {161},
  number = {3},
  pages = {147},
  doi = {10.3847/1538-3881/abd806}
}

@article{ballhausen2017,
  title = {Looking at {{A}} 0535+26 at Low Luminosities with {{{\emph{NuSTAR}}}}},
  author = {Ballhausen, Ralf and Pottschmidt, Katja and F{\"u}rst, Felix and Wilms, J{\"o}rn and Tomsick, John A. and Schwarm, Fritz-Walter and Stern, Daniel and Kretschmar, Peter and Caballero, Isabel and Harrison, Fiona A. and Boggs, Steven E. and Christensen, Finn E. and Craig, William W. and Hailey, Charles J. and Zhang, William W.},
  year = {2017},
  month = dec,
  journal = {A\&A},
  volume = {608},
  pages = {A105},
  doi = {10.1051/0004-6361/201730845}
}

@article{barthelmy2016,
  title = {Swift Detection of a Likely New {{SGR}}: {{SGR}} 0755-2933},
  shorttitle = {Swift Detection of a Likely New {{SGR}}},
  author = {Barthelmy, S. D. and D'Elia, V. and Gehrels, N. and Izzo, L. and Kennea, J. A. and Krimm, H. A. and Palmer, D. M. and Siegel, M. H. and Ukwatta, T. N.},
  year = {2016},
  month = mar,
  journal = {ATel},
  volume = {8831},
  langid = {english}
}

@article{basko1975,
  title = {Radiative Transfer in a Strong Magnetic Field and Accreting {{X-ray}} Pulsars.},
  author = {Basko, M. M. and Sunyaev, R. A.},
  year = {1975},
  month = sep,
  journal = {A\&A},
  volume = {42},
  pages = {311--321}
}

@article{basko1976,
  title = {The Limiting Luminosity of Accreting Neutron Stars with Magnetic Fields.},
  author = {Basko, M. M. and Sunyaev, R. A.},
  year = {1976},
  month = may,
  journal = {MNRAS},
  volume = {175},
  pages = {395--417},
  publisher = {OUP},
  doi = {10.1093/mnras/175.2.395}
}

@article{becker2012,
  title = {Spectral Formation in Accreting {{X-ray}} Pulsars: Bimodal Variation of the Cyclotron Energy with Luminosity},
  author = {Becker, P. A. and Klochkov, D. and Sch{\"o}nherr, G. and Nishimura, O. and Ferrigno, C. and Caballero, I. and Kretschmar, P. and Wolff, M. T. and Wilms, J. and Staubert, R.},
  year = {2012},
  month = aug,
  journal = {A\&A},
  volume = {544},
  number = {A123},
  pages = {A123},
  doi = {10.1051/0004-6361/201219065}
}

@article{brumback2018,
  title = {A {{Possible Phase-dependent Absorption Feature}} in the {{Transient X-Ray Pulsar SAX J2103}}.5+4545},
  author = {Brumback, M. C. and Hickox, R. C. and F{\"u}rst, F. S. and Pottschmidt, K. and Hemphill, P. and Tomsick, J. A. and Wilms, J. and Ballhausen, R.},
  year = {2018},
  month = jan,
  journal = {ApJ},
  volume = {852},
  number = {2},
  pages = {132},
  doi = {10.3847/1538-4357/aa9e91},
  langid = {english}
}

@article{burrows2005,
  title = {The {{Swift X-Ray Telescope}}},
  author = {Burrows, David N. and Hill, J. E. and Nousek, J. A. and Kennea, J. A. and Wells, A. and Osborne, J. P. and Abbey, A. F. and Beardmore, A. and Mukerjee, K. and Short, A. D. T. and Chincarini, G. and Campana, S. and Citterio, O. and Moretti, A. and Pagani, C. and Tagliaferri, G. and Giommi, P. and Capalbi, M. and Tamburelli, F. and Angelini, L. and Cusumano, G. and Br{\"a}uninger, H. W. and Burkert, W. and Hartner, G. D.},
  year = {2005},
  month = oct,
  journal = {Space Sci. Rev.},
  volume = {120},
  pages = {165--195},
  doi = {10.1007/s11214-005-5097-2}
}

@article{caballero2007,
  title = {A 0535+26 in the {{August}}/{{September}} 2005 Outburst Observed by {{RXTE}} and {{INTEGRAL}}},
  author = {Caballero, I. and Kretschmar, P. and Santangelo, A. and Staubert, R. and Klochkov, D. and Camero, A. and Ferrigno, C. and Finger, M. H. and Kreykenbohm, I. and McBride, V. A. and Pottschmidt, K. and Rothschild, R. E. and Sch{\"o}nherr, G. and Segreto, A. and Suchy, S. and Wilms, J. and Wilson, C. A.},
  year = {2007},
  month = apr,
  journal = {A\&A},
  volume = {465},
  number = {2},
  pages = {L21},
  doi = {10.1051/0004-6361:20067032},
  langid = {english}
}

@article{campana2002,
  title = {The {{Quiescent X-Ray Emission}} of {{Three Transient X-Ray Pulsars}}},
  author = {Campana, S. and Stella, L. and Israel, G. L. and Moretti, A. and Parmar, A. N. and Orlandini, M.},
  year = {2002},
  month = nov,
  journal = {ApJ},
  volume = {580},
  pages = {389--393},
  publisher = {IOP},
  doi = {10.1086/343074}
}

@inproceedings{chaty2011,
  title = {Nature, {{Formation}}, and {{Evolution}} of {{High Mass X-Ray Binaries}}},
  booktitle = {Evolution of Compact Binaries},
  author = {Chaty, S.},
  editor = {Schmidtobreick, Linda and Schreiber, Matthias R. and Tappert, Claus},
  year = {2011},
  month = sep,
  series = {{{ASP Conference Proceedings}}},
  volume = {447},
  pages = {29},
  publisher = {ASPCS},
  doi = {10.48550/arXiv.1107.0231}
}

@article{chen2021,
  title = {Relation of {{Cyclotron Resonant Energy}} and {{Luminosity}} in a {{Strongly Magnetized Neutron Star GRO J1008-57 Observed}} by {{Insight-HXMT}}},
  author = {Chen, X. and Wang, W. and Tang, Y. M. and Ding, Y. Z. and Tuo, Y. L. and Mushtukov, A. A. and Nishimura, O. and Zhang, S. N. and Ge, M. Y. and Song, L. M. and Lu, F. J. and Zhang, S. and Qu, J. L.},
  year = {2021},
  month = sep,
  journal = {ApJ},
  volume = {919},
  pages = {33},
  publisher = {IOP},
  doi = {10.3847/1538-4357/ac1268}
}

@article{coburn2001,
  title = {Discovery of a Cyclotron Resonant Scattering Feature in the Rossi {{X-ray}} Timing Explorer Spectrum of {{4U}} 0352+309 ({{X}} Persei)},
  author = {Coburn, W. and Heindl, W. A. and Gruber, D. E. and Rothschild, R. E. and Staubert, R. and Wilms, J. and Kreykenbohm, I.},
  year = {2001},
  month = may,
  journal = {ApJ},
  volume = {552},
  number = {2},
  eprint = {astro-ph/0101110},
  pages = {738--747},
  doi = {10.1086/320565},
  archiveprefix = {arXiv}
}

@article{coburn2002,
  title = {Magnetic {{Fields}} of {{Accreting X}}-{{Ray Pulsars}} with the {{{\emph{Rossi X}}}}{\emph{-}}{{{\emph{Ray Timing Explorer}}}}},
  author = {Coburn, W. and Heindl, W. A. and Rothschild, R. E. and Gruber, D. E. and Kreykenbohm, I. and Wilms, J. and Kretschmar, P. and Staubert, R.},
  year = {2002},
  month = nov,
  journal = {ApJ},
  volume = {580},
  number = {1},
  pages = {394--412},
  doi = {10.1086/343033},
  langid = {english}
}

@article{delgado-marti2001,
  title = {The {{Orbit}} of {{X Persei}} and {{Its Neutron Star Companion}}},
  author = {{Delgado-Mart{\'i}}, Hugo and Levine, Alan M. and Pfahl, Eric and Rappaport, Saul A.},
  year = {2001},
  month = jan,
  journal = {ApJ},
  volume = {546},
  pages = {455--468},
  publisher = {IOP},
  doi = {10.1086/318236}
}

@article{denherder2001,
  title = {The {{Reflection Grating Spectrometer}} on Board {{XMM-Newton}}},
  author = {{den Herder}, J. W. and Brinkman, A. C. and Kahn, S. M. and {Branduardi-Raymont}, G. and Thomsen, K. and Aarts, H. and Audard, M. and Bixler, J. V. and {den Boggende}, A. J. and Cottam, J. and Decker, T. and Dubbeldam, L. and Erd, C. and Goulooze, H. and G{\"u}del, M. and Guttridge, P. and Hailey, C. J. and Janabi, K. Al and Kaastra, J. S. and {de Korte}, P. A. J. and {van Leeuwen}, B. J. and Mauche, C. and McCalden, A. J. and Mewe, R. and Naber, A. and Paerels, F. B. and Peterson, J. R. and Rasmussen, A. P. and Rees, K. and Sakelliou, I. and Sako, M. and Spodek, J. and Stern, M. and Tamura, T. and Tandy, J. and {de Vries}, C. P. and Welch, S. and Zehnder, A.},
  year = {2001},
  month = jan,
  journal = {A\&A},
  volume = {365},
  pages = {L7-L17},
  doi = {10.1051/0004-6361:20000058}
}

@article{disalvo1998,
  title = {The {{Two-Component X-Ray Broadband Spectrum}} of {{X Persei Observed}} by {{BeppoSAX}}},
  author = {Di Salvo, T. and Burderi, L. and Robba, N. R. and Guainazzi, M.},
  year = {1998},
  month = dec,
  journal = {ApJ},
  volume = {509},
  pages = {897--903},
  publisher = {IOP},
  doi = {10.1086/306525}
}

@article{doroshenko2012,
  title = {The Hard {{X-ray}} Emission of {{X Persei}}},
  author = {Doroshenko, V. and Santangelo, A. and Kreykenbohm, I. and Doroshenko, R.},
  year = {2012},
  month = apr,
  journal = {A\&A},
  volume = {540},
  pages = {L1},
  doi = {10.1051/0004-6361/201218878}
}

@article{doroshenko2021,
  title = {{{SGR}} 0755-2933: A New High-Mass {{X-ray}} Binary with the Wrong Name},
  shorttitle = {{{SGR}} 0755-2933},
  author = {Doroshenko, V. and Santangelo, A. and Tsygankov, S. S. and Ji, L.},
  year = {2021},
  month = mar,
  journal = {A\&A},
  volume = {647},
  pages = {A165},
  doi = {10.1051/0004-6361/202039785},
  copyright = {https://www.edpsciences.org/en/authors/copyright-and-licensing},
  langid = {english}
}

@article{doroshenko2022,
  title = {{{SRGA J124404}}.1-632232/{{SRGU J124403}}.8-632231: {{New X-ray}} Pulsar Discovered in the All-Sky Survey by the {{SRG}}},
  shorttitle = {{{SRGA J124404}}.1-632232/{{SRGU J124403}}.8-632231},
  author = {Doroshenko, V. and Staubert, R. and Maitra, C. and Rau, A. and Haberl, F. and Santangelo, A. and Schwope, A. and Wilms, J. and Buckley, D. A. H. and Semena, A. and Mereminskiy, I. and Lutovinov, A. and Gromadzki, M. and Townsend, L. J. and Monageng, I. M.},
  year = {2022},
  month = may,
  journal = {A\&A},
  volume = {661},
  pages = {A21},
  doi = {10.1051/0004-6361/202141147}
}

@inproceedings{finger1994,
  title = {Hard {{X-Ray Observations}} of {{A}} 0535+262},
  booktitle = {The {{Evolution}} of {{X-ray Binaries}}},
  author = {Finger, M. H. and Cominsky, L. R. and Wilson, R. B. and Harmon, B. A. and Fishman, G. J.},
  editor = {Holt, Steve and Day, Charles S.},
  year = {1994},
  month = jan,
  series = {{{AIP Conference Proceedings}}},
  volume = {308},
  pages = {459},
  publisher = {AIP},
  address = {New York, NY},
  doi = {10.1063/1.46032},
  langid = {english}
}

@article{furst2015,
  title = {Distorted {{Cyclotron Line Profile}} in {{Cep X-4}} as {{Observed}} by {{NuSTAR}}},
  author = {F{\"u}rst, F. and Pottschmidt, K. and Miyasaka, H. and Bhalerao, V. and Bachetti, M. and Boggs, S. E. and Christensen, F. E. and Craig, W. W. and Grinberg, V. and Hailey, C. J. and Harrison, F. A. and Kennea, J. A. and Rahoui, F. and Stern, D. and Tendulkar, S. P. and Tomsick, J. A. and Walton, D. J. and Wilms, J. and Zhang, W. W.},
  year = {2015},
  month = jun,
  journal = {ApJ},
  volume = {806},
  pages = {L24},
  publisher = {IOP},
  doi = {10.1088/2041-8205/806/2/L24}
}

@article{furst2018,
  title = {Multiple Cyclotron Line-Forming Regions in {{GX}} 301-2},
  author = {F{\"u}rst, F. and Falkner, S. and {Marcu-Cheatham}, D. and Grefenstette, B. and Tomsick, J. and Pottschmidt, K. and Walton, D. J. and Natalucci, L. and Kretschmar, P.},
  year = {2018},
  month = dec,
  journal = {A\&A},
  volume = {620},
  eprint = {1809.05691},
  pages = {A153},
  doi = {10.1051/0004-6361/201732132},
  archiveprefix = {arXiv}
}

@article{gaiacollaboration2016,
  title = {The {{{\emph{Gaia}}}} Mission},
  author = {{Gaia Collaboration}},
  year = {2016},
  month = nov,
  journal = {A\&A},
  volume = {595},
  pages = {A1},
  doi = {10.1051/0004-6361/201629272}
}

@article{gaiacollaboration2023,
  title = {Gaia {{Data Release}} 3. {{Summary}} of the Content and Survey Properties},
  author = {{Gaia Collaboration}},
  year = {2023},
  month = jun,
  journal = {A\&A},
  volume = {674},
  pages = {A1},
  doi = {10.1051/0004-6361/202243940}
}

@article{gehrels2004,
  title = {The {{{\emph{Swift}}}} {{Gamma}}-{{Ray Burst Mission}}},
  author = {Gehrels, N. and Chincarini, G. and Giommi, P. and Mason, K. O. and Nousek, J. A. and Wells, A. A. and White, N. E. and Barthelmy, S. D. and Burrows, D. N. and Cominsky, L. R. and Hurley, K. C. and Marshall, F. E. and Meszaros, P. and Roming, P. W. A. and Angelini, L. and Barbier, L. M. and Belloni, T. and Campana, S. and Caraveo, P. A. and Chester, M. M. and Citterio, O. and Cline, T. L. and Cropper, M. S. and Cummings, J. R. and Dean, A. J. and Feigelson, E. D. and Fenimore, E. E. and Frail, D. A. and Fruchter, A. S. and Garmire, G. P. and Gendreau, K. and Ghisellini, G. and Greiner, J. and Hill, J. E. and Hunsberger, S. D. and Krimm, H. A. and Kulkarni, S. R. and Kumar, P. and Lebrun, F. and Lloyd-Ronning, N. M. and Markwardt, C. B. and Mattson, B. J. and Mushotzky, R. F. and Norris, J. P. and Osborne, J. and Paczynski, B. and Palmer, D. M. and Park, H.-S. and Parsons, A. M. and Paul, J. and Rees, M. J. and Reynolds, C. S. and Rhoads, J. E. and Sasseen, T. P. and Schaefer, B. E. and Short, A. T. and Smale, A. P. and Smith, I. A. and Stella, L. and Tagliaferri, G. and Takahashi, T. and Tashiro, M. and Townsley, L. K. and Tueller, J. and Turner, M. J. L. and Vietri, M. and Voges, W. and Ward, M. J. and Willingale, R. and Zerbi, F. M. and Zhang, W. W.},
  year = {2004},
  month = aug,
  journal = {ApJ},
  volume = {611},
  number = {2},
  pages = {1005--1020},
  doi = {10.1086/422091},
  langid = {english}
}

@inproceedings{gendreau2012,
  title = {The {{Neutron}} Star {{Interior Composition ExploreR}} ({{NICER}}): An {{Explorer}} Mission of Opportunity for Soft x-Ray Timing Spectroscopy},
  shorttitle = {The {{Neutron}} Star {{Interior Composition ExploreR}} ({{NICER}})},
  booktitle = {Space {{Telescopes}} and {{Instrumentation}} 2012: {{Ultraviolet}} to {{Gamma Ray}}},
  author = {Gendreau, Keith C. and Arzoumanian, Zaven and Okajima, Takashi},
  editor = {Takahashi, Tadayuki and Murray, Stephen S. and {den Herder}, Jan-Willem A.},
  year = {2012},
  month = sep,
  volume = {8443},
  pages = {322--329},
  publisher = {SPIE},
  doi = {10.1117/12.926396}
}

@inproceedings{gendreau2016,
  title = {The {{Neutron}} Star {{Interior Composition Explorer}} ({{NICER}}): Design and Development},
  shorttitle = {The {{Neutron}} Star {{Interior Composition Explorer}} ({{NICER}})},
  booktitle = {Space {{Telescopes}} and {{Instrumentation}} 2016: {{Ultraviolet}} to {{Gamma Ray}}},
  author = {Gendreau, Keith C. and Arzoumanian, Zaven and Adkins, Phillip W. and Albert, Cheryl L. and Anders, John F. and Aylward, Andrew T. and Baker, Charles L. and Balsamo, Erin R. and Bamford, William A. and Benegalrao, Suyog S. and Berry, Daniel L. and Bhalwani, Shiraz and Black, J. Kevin and Blaurock, Carl and Bronke, Ginger M. and Brown, Gary L. and Budinoff, Jason G. and Cantwell, Jeffrey D. and Cazeau, Thoniel and Chen, Philip T. and Clement, Thomas G. and Colangelo, Andrew T. and Coleman, Jerry S. and Coopersmith, Jonathan D. and Dehaven, William E. and Doty, John P. and Egan, Mark D. and Enoto, Teruaki and Fan, Terry W.-M. and Ferro, Deneen M. and Foster, Richard and Galassi, Nicholas M. and Gallo, Luis D. and Green, Chris M. and Grosh, Dave and Ha, Kong Q. and Hasouneh, Monther A. and Heefner, Kristofer B. and Hestnes, Phyllis and Hoge, Lisa J. and Jacobs, Tawanda M. and J{\o}rgensen, John L. and Kaiser, Michael A. and Kellogg, James W. and Kenyon, Steven J. and Koenecke, Richard G. and Kozon, Robert P. and LaMarr, Beverly and Lambertson, Mike D. and Larson, Anne M. and Lentine, Steven and Lewis, Jesse H. and Lilly, Michael G. and Liu, Kuochia Alice and Malonis, Andrew and Manthripragada, Sridhar S. and Markwardt, Craig B. and Matonak, Bryan D. and Mcginnis, Isaac E. and Miller, Roger L. and Mitchell, Alissa L. and Mitchell, Jason W. and Mohammed, Jelila S. and Monroe, Charles A. and de Garcia, Kristina M. Montt and Mul{\'e}, Peter D. and Nagao, Louis T. and Ngo, Son N. and Norris, Eric D. and Norwood, Dwight A. and Novotka, Joseph and Okajima, Takashi and Olsen, Lawrence G. and Onyeachu, Chimaobi O. and Orosco, Henry Y. and Peterson, Jacqualine R. and Pevear, Kristina N. and Pham, Karen K. and Pollard, Sue E. and Pope, John S. and Powers, Daniel F. and Powers, Charles E. and Price, Samuel R. and Prigozhin, Gregory Y. and Ramirez, Julian B. and Reid, Winston J. and Remillard, Ronald A. and Rogstad, Eric M. and Rosecrans, Glenn P. and Rowe, John N. and Sager, Jennifer A. and Sanders, Claude A. and Savadkin, Bruce and Saylor, Maxine R. and Schaeffer, Alexander F. and Schweiss, Nancy S. and Semper, Sean R. and Serlemitsos, Peter J. and Shackelford, Larry V. and Soong, Yang and Struebel, Jonathan and Vezie, Michael L. and Villasenor, Joel S. and Winternitz, Luke B. and Wofford, George I. and Wright, Michael R. and Yang, Mike Y. and Yu, Wayne H.},
  editor = {{den Herder}, Jan-Willem A. and Takahashi, Tadayuki and Bautz, Marshall},
  year = {2016},
  month = jul,
  volume = {9905},
  pages = {420--435},
  publisher = {SPIE},
  doi = {10.1117/12.2231304}
}

@article{ghimiray2024,
  title = {Low-Luminosity Accretion of {{Be}}/{{X-ray}} Pulsar {{MAXI J1409-619}}},
  author = {Ghimiray, Monika and Sharma, Pankaj and Subba, Nishika},
  year = {2024},
  month = jul,
  journal = {MNRAS},
  volume = {531},
  pages = {3386--3390},
  publisher = {OUP},
  doi = {10.1093/mnras/stae1369}
}

@article{ghising2023a,
  title = {Low-Luminosity Observation of {{BeXRB}} Source {{IGR J21347}}{$<$}inline-Formula Id="{{IEq1}}"{$><$}mml:Math{$><$}mml:Mo{$>$}+{$<$}/Mml:Mo{$><$}/Mml:Math{$><$}/Inline-Formula{$>$}4737},
  shorttitle = {Low-Luminosity Observation of {{BeXRB}} Source {{IGR J21347}}{$<$}inline-Formula Id="{{IEq1}}"{$>$}},
  author = {Ghising, Manoj and Tamang, Ruchi and Rai, Binay and Tobrej, Mohammed and Paul, Bikash Chandra},
  year = {2023},
  month = dec,
  journal = {J. Astrophys. Astr.},
  volume = {44},
  number = {2},
  pages = {94},
  doi = {10.1007/s12036-023-09986-0},
  langid = {english}
}

@article{ghosh1979,
  title = {Accretion by Rotating Magnetic Neutron Stars. {{III}}. {{Accretion}} Torques and Period Changes in Pulsating {{X-ray}} Sources.},
  author = {Ghosh, P. and Lamb, F. K.},
  year = {1979},
  month = nov,
  journal = {ApJ},
  volume = {234},
  pages = {296--316},
  publisher = {IOP},
  doi = {10.1086/157498}
}

@article{giacconi1972,
  title = {The {{Uhuru}} Catalog of {{X-ray}} Sources.},
  author = {Giacconi, R. and Murray, S. and Gursky, H. and Kellogg, E. and Schreier, E. and Tananbaum, H.},
  year = {1972},
  month = dec,
  journal = {ApJ},
  volume = {178},
  pages = {281--308},
  publisher = {IOP},
  doi = {10.1086/151790}
}

@article{gorban2022,
  title = {Changes in the {{Nature}} of the {{Spectral Continuum}} and {{Stability}} of the {{Cyclotron Line}} in the {{X-ray Pulsar GRO J2058}}+42},
  author = {Gorban, A. S. and Molkov, S. V. and Tsygankov, S. S. and Mushtukov, A. A. and Lutovinov, A. A.},
  year = {2022},
  month = apr,
  journal = {Astron. Lett.},
  volume = {48},
  number = {4},
  pages = {256--266},
  doi = {10.1134/S1063773722040028},
  langid = {english}
}

@article{gorban2022a,
  title = {Study of the {{X-ray Pulsar IGR J21343}}+4738 {{Based}} on {{NuSTAR}}, {{Swift}}, and {{SRG Data}}},
  author = {Gorban, A. S. and Molkov, S. V. and Lutovinov, A. A. and Semena, A. N.},
  year = {2022},
  month = dec,
  journal = {Astron. Lett.},
  volume = {48},
  number = {12},
  pages = {798--805},
  doi = {10.1134/S106377372211007X},
  langid = {english}
}

@article{harding1984,
  title = {Self-Consistent Models for {{Coulomb-heated X-ray}} Pulsar Atmospheres.},
  author = {Harding, A. K. and Meszaros, P. and Kirk, J. G. and Galloway, D. J.},
  year = {1984},
  month = mar,
  journal = {ApJ},
  volume = {278},
  pages = {369--381},
  publisher = {IOP},
  doi = {10.1086/161801}
}

@article{harding2006,
  title = {Physics of Strongly Magnetized Neutron Stars},
  author = {Harding, Alice K. and Lai, Dong},
  year = {2006},
  month = sep,
  journal = {RPPh},
  volume = {69},
  pages = {2631--2708},
  publisher = {IOP},
  doi = {10.1088/0034-4885/69/9/R03}
}

@article{harrison2013,
  title = {{{THE}} {{{\emph{NUCLEAR SPECTROSCOPIC TELESCOPE ARRAY}}}} ( {{{\emph{NuSTAR}}}} ) {{HIGH-ENERGY X-RAY MISSION}}},
  author = {Harrison, Fiona A. and Craig, William W. and Christensen, Finn E. and Hailey, Charles J. and Zhang, William W. and Boggs, Steven E. and Stern, Daniel and Cook, W. Rick and Forster, Karl and Giommi, Paolo and Grefenstette, Brian W. and Kim, Yunjin and Kitaguchi, Takao and Koglin, Jason E. and Madsen, Kristin K. and Mao, Peter H. and Miyasaka, Hiromasa and Mori, Kaya and Perri, Matteo and Pivovaroff, Michael J. and Puccetti, Simonetta and Rana, Vikram R. and Westergaard, Niels J. and Willis, Jason and Zoglauer, Andreas and An, Hongjun and Bachetti, Matteo and Barri{\`e}re, Nicolas M. and Bellm, Eric C. and Bhalerao, Varun and Brejnholt, Nicolai F. and Fuerst, Felix and Liebe, Carl C. and Markwardt, Craig B. and Nynka, Melania and Vogel, Julia K. and Walton, Dominic J. and Wik, Daniel R. and Alexander, David M. and Cominsky, Lynn R. and Hornschemeier, Ann E. and Hornstrup, Allan and Kaspi, Victoria M. and Madejski, Greg M. and Matt, Giorgio and Molendi, Silvano and Smith, David M. and Tomsick, John A. and Ajello, Marco and Ballantyne, David R. and Balokovi{\'c}, Mislav and Barret, Didier and Bauer, Franz E. and Blandford, Roger D. and Brandt, W. Niel and Brenneman, Laura W. and Chiang, James and Chakrabarty, Deepto and Chenevez, Jerome and Comastri, Andrea and Dufour, Francois and Elvis, Martin and Fabian, Andrew C. and Farrah, Duncan and Fryer, Chris L. and Gotthelf, Eric V. and Grindlay, Jonathan E. and Helfand, David J. and Krivonos, Roman and Meier, David L. and Miller, Jon M. and Natalucci, Lorenzo and Ogle, Patrick and Ofek, Eran O. and Ptak, Andrew and Reynolds, Stephen P. and Rigby, Jane R. and Tagliaferri, Gianpiero and Thorsett, Stephen E. and Treister, Ezequiel and Urry, C. Megan},
  year = {2013},
  month = may,
  journal = {ApJ},
  volume = {770},
  number = {2},
  pages = {103},
  doi = {10.1088/0004-637X/770/2/103}
}

@inproceedings{harrison2017,
  title = {Searching for {{Magnetar SGR}} 0755-2933},
  booktitle = {{{AAS Meeting}} \#229, Id.431.04},
  author = {Harrison, Amanda and Lynch, Ryan},
  year = {2017},
  month = jan,
  publisher = {American Astronomical Society},
  address = {Grapevine, Texas},
  langid = {english}
}

@article{hi4picollaboration2016,
  title = {{{HI4PI}}: {{A}} Full-Sky {{H I}} Survey Based on {{EBHIS}} and {{GASS}}},
  shorttitle = {{{HI4PI}}},
  author = {{HI4PI Collaboration} and Ben Bekhti, N. and Fl{\"o}er, L. and Keller, R. and Kerp, J. and Lenz, D. and Winkel, B. and Bailin, J. and Calabretta, M. R. and Dedes, L. and Ford, H. A. and Gibson, B. K. and Haud, U. and Janowiecki, S. and Kalberla, P. M. W. and Lockman, F. J. and {McClure-Griffiths}, N. M. and Murphy, T. and Nakanishi, H. and Pisano, D. J. and {Staveley-Smith}, L.},
  year = {2016},
  month = oct,
  journal = {A\&A},
  volume = {594},
  pages = {A116},
  doi = {10.1051/0004-6361/201629178}
}

@inproceedings{houck2000,
  title = {{{ISIS}}: {{An}} Interactive Spectral Interpretation System for High Resolution {{X-ray}} Spectroscopy},
  booktitle = {Astronomical Data Analysis Software and Systems {{IX}}},
  author = {Houck, J. C. and Denicola, L. A.},
  editor = {Manset, Nadine and Veillet, Christian and Crabtree, Dennis},
  year = {2000},
  month = jan,
  series = {{{ASP Conference Series}}},
  volume = {216},
  pages = {591}
}

@article{illarionov1990,
  title = {A {{Spin-Down Mechanism}} for {{Accreting Neutron Stars}}},
  author = {Illarionov, A. F. and Kompaneets, D. A.},
  year = {1990},
  month = nov,
  journal = {MNRAS},
  volume = {247},
  pages = {219},
  publisher = {OUP}
}

@article{jansen2001,
  title = {{{XMM-Newton}} Observatory. {{I}}. {{The}} Spacecraft and Operations},
  author = {Jansen, F. and Lumb, D. and Altieri, B. and Clavel, J. and Ehle, M. and Erd, C. and Gabriel, C. and Guainazzi, M. and Gondoin, P. and Much, R. and Munoz, R. and Santos, M. and Schartel, N. and Texier, D. and Vacanti, G.},
  year = {2001},
  month = jan,
  journal = {A\&A},
  volume = {365},
  pages = {L1-L6},
  doi = {10.1051/0004-6361:20000036}
}

@article{kreykenbohm2004,
  title = {The Variable Cyclotron Line in {{GX}} 301-2},
  author = {Kreykenbohm, Ingo and Wilms, Joern and Coburn, Wayne and Kuster, Markus and Rothschild, Richard E. and Heindl, William A. and Kretschmar, Peter and Staubert, Ruediger},
  year = {2004},
  month = dec,
  journal = {A\&A},
  volume = {427},
  number = {3},
  eprint = {astro-ph/0409015},
  pages = {975--986},
  doi = {10.1051/0004-6361:20035836},
  archiveprefix = {arXiv}
}

@article{krivonos2015,
  title = {{{NuSTAR Discovery}} of an {{Unusually Steady Long-term Spin-up}} of the {{Be Binary 2RXP J130159}}.6-635806},
  author = {Krivonos, Roman A. and Tsygankov, Sergey S. and Lutovinov, Alexander A. and Tomsick, John A. and Chakrabarty, Deepto and Bachetti, Matteo and Boggs, Steven E. and Chernyakova, Masha and Christensen, Finn E. and Craig, William W. and F{\"u}rst, Felix and Hailey, Charles J. and Harrison, Fiona A. and Lansbury, George B. and Rahoui, Farid and Stern, Daniel and Zhang, William W.},
  year = {2015},
  month = aug,
  journal = {ApJ},
  volume = {809},
  pages = {140},
  publisher = {IOP},
  doi = {10.1088/0004-637X/809/2/140}
}

@article{kuhnel2013,
  title = {{{GRO J1008}}-57: An (Almost) Predictable Transient {{X-ray}} Binary},
  shorttitle = {{{GRO J1008}}-57},
  author = {K{\"u}hnel, M. and M{\"u}ller, S. and Kreykenbohm, I. and F{\"u}rst, F. and Pottschmidt, K. and Rothschild, R. E. and Caballero, I. and Grinberg, V. and Sch{\"o}nherr, G. and Shrader, C. and Klochkov, D. and Staubert, R. and Ferrigno, C. and Torrej{\'o}n, J.-M. and {Mart{\'i}nez-N{\'u}{\~n}ez}, S. and Wilms, J.},
  year = {2013},
  month = jul,
  journal = {A\&A},
  volume = {555},
  pages = {A95},
  doi = {10.1051/0004-6361/201321203},
  langid = {english}
}

@article{lamb1973,
  title = {A {{Model}} for {{Compact X-Ray Sources}}: {{Accretion}} by {{Rotating Magnetic Stars}}},
  shorttitle = {A {{Model}} for {{Compact X-Ray Sources}}},
  author = {Lamb, F. K. and Pethick, C. J. and Pines, D.},
  year = {1973},
  month = aug,
  journal = {ApJ},
  volume = {184},
  pages = {271},
  doi = {10.1086/152325},
  langid = {english}
}

@article{ludlam2022,
  title = {{{StrayCats}}. {{II}}. {{An Updated Catalog}} of {{NuSTAR Stray Light Observations}}},
  author = {Ludlam, R. M. and Grefenstette, B. W. and Brumback, M. C. and Tomsick, J. A. and Buisson, D. J. K. and Coughenour, B. M. and Mastroserio, G. and Wik, D. and Krivonos, R. and Jaodand, A. D. and Madsen, K. K.},
  year = {2022},
  month = jul,
  journal = {ApJ},
  volume = {934},
  number = {1},
  pages = {59},
  doi = {10.3847/1538-4357/ac7b27},
  langid = {english}
}

@article{lutovinov2012,
  title = {Strong Outburst Activity of the {{X-ray}} Pulsar {{X Persei}} during 2001-2011},
  author = {Lutovinov, A. and Tsygankov, S. and Chernyakova, M.},
  year = {2012},
  month = jun,
  journal = {MNRAS},
  volume = {423},
  pages = {1978--1984},
  publisher = {OUP},
  doi = {10.1111/j.1365-2966.2012.21036.x}
}

@article{lutovinov2017,
  title = {{{NuSTAR}} Observations of the Supergiant {{X-ray}} Pulsar {{IGR J18027-2016}}: Accretion from the Stellar Wind and Possible Cyclotron Absorption Line},
  shorttitle = {{{NuSTAR}} Observations of the Supergiant {{X-ray}} Pulsar {{IGR J18027-2016}}},
  author = {Lutovinov, Alexander A. and Tsygankov, Sergey S. and Postnov, Konstantin A. and Krivonos, Roman A. and Molkov, Sergey V. and Tomsick, John A.},
  year = {2017},
  month = apr,
  journal = {MNRAS},
  volume = {466},
  number = {1},
  pages = {593--599},
  doi = {10.1093/mnras/stw3058},
  langid = {english}
}

@article{lutovinov2021,
  title = {{{SRG}}/{{ART-XC}} and {{NuSTAR Observations}} of the {{X-Ray}} Pulsar {{GRO J1008-57}} in the {{Lowest Luminosity State}}},
  author = {Lutovinov, A. and Tsygankov, S. and Molkov, S. and Doroshenko, V. and Mushtukov, A. and Arefiev, V. and Lapshov, I. and Tkachenko, A. and Pavlinsky, M.},
  year = {2021},
  month = may,
  journal = {ApJ},
  volume = {912},
  pages = {17},
  publisher = {IOP},
  doi = {10.3847/1538-4357/abec43}
}

@article{makishima1999,
  title = {Cyclotron {{Resonance Effects}} in {{Two Binary X-Ray Pulsars}} and the {{Evolution}} of {{Neutron Star Magnetic Fields}}},
  author = {Makishima, K. and Mihara, T. and Nagase, F. and Tanaka, Y.},
  year = {1999},
  month = nov,
  journal = {ApJ},
  volume = {525},
  number = {2},
  pages = {978},
  doi = {10.1086/307912},
  langid = {english}
}

@article{malacaria2015,
  title = {Luminosity-Dependent Spectral and Timing Properties of the Accreting Pulsar {{GX}} 304-1 Measured with {{INTEGRAL}}},
  author = {Malacaria, C. and Klochkov, D. and Santangelo, A. and Staubert, R.},
  year = {2015},
  month = sep,
  journal = {A\&A},
  volume = {581},
  pages = {A121},
  doi = {10.1051/0004-6361/201526417}
}

@article{malacaria2026,
  title = {Double-Hump Spectrum, Pulse Profile Dip, and Pulsed Fraction Spectra from the Low-Accretion Regime in the {{X-ray}} Pulsar {{MAXI J0655-013}}},
  author = {Malacaria, C. and Pike, S. N. and D'A{\`i}, A. and Israel, G. L. and Ducci, L. and Rothschild, R. E. and Stella, L. and Amato, R. and Ambrosi, E. and Coley, J. B. and F{\"u}rst, F. and Imbrogno, M. and Kretschmar, P. and Maniadakis, D. K. and Papitto, A. and Pradhan, P. and Rouco Escorial, A. and Simongini, A. and Stierhof, J. and West, B. F. and Zalot, N.},
  year = {2026},
  month = feb,
  journal = {A\&A},
  volume = {706},
  pages = {A321},
  publisher = {EDP},
  doi = {10.1051/0004-6361/202557862}
}

@inproceedings{markwardt2024,
  title = {{{NICER SCORPEON Background Model}}},
  booktitle = {Bulletin of the {{American Astronomical Society}}},
  author = {Markwardt, Craig and Arzoumanian, Zaven and Gendreau, Keith and Hare, Jeremy and {NICER Team}},
  editor = {Vishniac, Ethan and Muench, August},
  year = {2024},
  month = may,
  volume = {56},
  publisher = {AAS Publishing},
  address = {Washington, DC},
  langid = {english}
}

@article{mcbride2007,
  title = {On the Cyclotron Line in {{Cepheus X-4}}},
  author = {McBride, V. A. and Wilms, J. and Kreykenbohm, I. and Coe, M. J. and Rothschild, R. E. and Kretschmar, P. and Pottschmidt, K. and Fisher, J. and Hamson, T.},
  year = {2007},
  month = aug,
  journal = {A\&A},
  volume = {470},
  number = {3},
  pages = {1065--1070},
  doi = {10.1051/0004-6361:20077238},
  langid = {english}
}

@article{merloni2024,
  title = {The {{SRG}}/{{eROSITA}} All-Sky Survey. {{First X-ray}} Catalogues and Data Release of the Western {{Galactic}} Hemisphere},
  author = {Merloni, A. and Lamer, G. and Liu, T. and {Ramos-Ceja}, M. E. and Brunner, H. and Bulbul, E. and Dennerl, K. and Doroshenko, V. and Freyberg, M. J. and Friedrich, S. and Gatuzz, E. and Georgakakis, A. and Haberl, F. and Igo, Z. and Kreykenbohm, I. and Liu, A. and Maitra, C. and Malyali, A. and Mayer, M. G. F. and Nandra, K. and Predehl, P. and Robrade, J. and Salvato, M. and Sanders, J. S. and Stewart, I. and {Tub{\'i}n-Arenas}, D. and Weber, P. and Wilms, J. and Arcodia, R. and Artis, E. and Aschersleben, J. and Avakyan, A. and Aydar, C. and Bahar, Y. E. and Balzer, F. and Becker, W. and Berger, K. and Boller, T. and Bornemann, W. and Br{\"u}ggen, M. and Brusa, M. and Buchner, J. and Burwitz, V. and Camilloni, F. and Clerc, N. and Comparat, J. and Coutinho, D. and Czesla, S. and Dannhauer, S. M. and Dauner, L. and Dauser, T. and Dietl, J. and Dolag, K. and Dwelly, T. and Egg, K. and Ehl, E. and Freund, S. and Friedrich, P. and Gaida, R. and Garrel, C. and Ghirardini, V. and Gokus, A. and Gr{\"u}nwald, G. and Grandis, S. and Grotova, I. and Gruen, D. and Gueguen, A. and H{\"a}mmerich, S. and Hamaus, N. and Hasinger, G. and Haubner, K. and Homan, D. and Ider Chitham, J. and Joseph, W. M. and Joyce, A. and K{\"o}nig, O. and Kaltenbrunner, D. M. and Khokhriakova, A. and Kink, W. and Kirsch, C. and Kluge, M. and Knies, J. and Krippendorf, S. and Krumpe, M. and Kurpas, J. and Li, P. and Liu, Z. and Locatelli, N. and Lorenz, M. and M{\"u}ller, S. and Magaudda, E. and Mannes, C. and McCall, H. and Meidinger, N. and Michailidis, M. and Migkas, K. and {Mu{\~n}oz-Giraldo}, D. and Musiimenta, B. and {Nguyen-Dang}, N. T. and Ni, Q. and Olechowska, A. and Ota, N. and Pacaud, F. and Pasini, T. and Perinati, E. and Pires, A. M. and Pommranz, C. and Ponti, G. and Poppenhaeger, K. and P{\"u}hlhofer, G. and Rau, A. and Reh, M. and Reiprich, T. H. and Roster, W. and Saeedi, S. and Santangelo, A. and Sasaki, M. and Schmitt, J. and Schneider, P. C. and Schrabback, T. and Schuster, N. and Schwope, A. and Seppi, R. and Serim, M. M. and Shreeram, S. and {Sokolova-Lapa}, E. and Starck, H. and Stelzer, B. and Stierhof, J. and Suleimanov, V. and Tenzer, C. and Traulsen, I. and Tr{\"u}mper, J. and Tsuge, K. and Urrutia, T. and Veronica, A. and Waddell, S. G. H. and Willer, R. and Wolf, J. and Yeung, M. C. H. and Zainab, A. and Zangrandi, F. and Zhang, X. and Zhang, Y. and Zheng, X.},
  year = {2024},
  month = feb,
  journal = {A\&A},
  volume = {682},
  pages = {A34},
  doi = {10.1051/0004-6361/202347165}
}

@article{meszaros1983,
  title = {Accreting {{X-ray}} Pulsar Atmospheres Heated by {{Coulomb}} Deceleration of Protons},
  author = {Meszaros, P. and Harding, A. K. and Kirk, J. G. and Galloway, D. J.},
  year = {1983},
  month = mar,
  journal = {ApJ},
  volume = {266},
  pages = {L33-L37},
  publisher = {IOP},
  doi = {10.1086/183973}
}

@article{meszaros1985a,
  title = {X-Ray Pulsar Models. {{II}}. {{Comptonized}} Spectra and Pulse Shapes.},
  author = {Meszaros, P. and Nagel, W.},
  year = {1985},
  month = dec,
  journal = {ApJ},
  volume = {299},
  pages = {138--153},
  doi = {10.1086/163687}
}

@article{miller1987,
  title = {Deceleration of {{Infalling Plasma}} in the {{Atmospheres}} of {{Accreting Neutron Stars}}. {{I}}. {{Isothermal Atmospheres}}},
  author = {Miller, G. S. and Salpeter, E. E. and Wasserman, I.},
  year = {1987},
  month = mar,
  journal = {ApJ},
  volume = {314},
  pages = {215},
  publisher = {IOP},
  doi = {10.1086/165051}
}

@article{miller1989,
  title = {The {{Deceleration}} of {{Infalling Plasma}} in {{Magnetized Neutron Star Atmospheres}}: {{Nonisothermal Atmospheres}}},
  shorttitle = {The {{Deceleration}} of {{Infalling Plasma}} in {{Magnetized Neutron Star Atmospheres}}},
  author = {Miller, Guy and Wasserman, Ira and Salpeter, Edwin E.},
  year = {1989},
  month = nov,
  journal = {ApJ},
  volume = {346},
  pages = {405},
  publisher = {IOP},
  doi = {10.1086/168020}
}

@article{mitsuda1984,
  title = {Energy {{Spectra}} of {{Low-Mass Binary X-Ray Sources Observed}} from {{Tenma}}},
  author = {Mitsuda, K. and Inoue, H. and Koyama, K. and Makishima, K. and Matsuoka, M. and Ogawara, Y. and Shibazaki, N. and Suzuki, K. and Tanaka, Y. and Hirano, T.},
  year = {1984},
  month = dec,
  journal = {PASJ},
  volume = {36},
  pages = {741--759},
  publisher = {OUP},
  doi = {10.1093/pasj/36.4.741}
}

@article{moffat1973,
  title = {On {{X-Persei}} as a Stellar {{X-ray}} Source: {{Comparison}} with Gamma {{Cassiopeiae}}.},
  shorttitle = {On {{X-Persei}} as a Stellar {{X-ray}} Source},
  author = {Moffat, A. F. J. and Haupt, W. and {Schmidt-Kaler}, T.},
  year = {1973},
  month = mar,
  journal = {A\&A},
  volume = {23},
  pages = {433}
}

@article{mroz2021,
  title = {Periodic Variability in the Archival {{OGLE}} Light Curve of {{X-ray}} Transient {{SRGA J124404}}.1-632232 ({{SRGE J124403}}.8-632231)},
  author = {Mroz, P. and Udalski, A.},
  year = {2021},
  month = jan,
  journal = {ATel},
  volume = {14361}
}

@article{mushtukov2015a,
  title = {The Critical Accretion Luminosity for Magnetized Neutron Stars},
  author = {Mushtukov, Alexander A. and Suleimanov, Valery F. and Tsygankov, Sergey S. and Poutanen, Juri},
  year = {2015},
  month = feb,
  journal = {MNRAS},
  volume = {447},
  pages = {1847--1856},
  publisher = {OUP},
  doi = {10.1093/mnras/stu2484}
}

@article{mushtukov2021,
  title = {Spectrum Formation in {{X-ray}} Pulsars at Very Low Mass Accretion Rate: {{Monte Carlo}} Approach},
  author = {Mushtukov, Alexander A. and Suleimanov, Valery F. and Tsygankov, Sergey S. and Portegies Zwart, Simon},
  year = {2021},
  month = may,
  journal = {MNRAS},
  volume = {503},
  number = {4},
  eprint = {2006.13596},
  primaryclass = {astro-ph.HE},
  pages = {5193--5203},
  doi = {10.1093/mnras/stab811},
  archiveprefix = {arXiv}
}

@article{mushtukov2023,
  title = {X-Ray Polarimetry of {{X-ray}} Pulsar {{X Persei}}: Another Orthogonal Rotator?},
  shorttitle = {X-Ray Polarimetry of {{X-ray}} Pulsar {{X Persei}}},
  author = {Mushtukov, A. A. and Tsygankov, S. S. and Poutanen, J. and Doroshenko, V. and Salganik, A. and Costa, E. and Marco, A. Di and Heyl, J. and Monaca, F. La and Lutovinov, A. A. and Mereminsky, I. A. and Papitto, A. and Semena, A. N. and Shtykovsky, A. E. and Suleimanov, V. F. and Forsblom, S. V. and {Gonz{\'a}lez-Caniulef}, D. and Malacaria, C. and Sunyaev, R. A. and Agudo, I. and Antonelli, L. A. and Bachetti, M. and Baldini, L. and Baumgartner, W. H. and Bellazzini, R. and Bianchi, S. and Bongiorno, S. D. and Bonino, R. and Brez, A. and Bucciantini, N. and Capitanio, F. and Castellano, S. and Cavazzuti, E. and Chen, C. -T. and Ciprini, S. and De Rosa, A. and Del Monte, E. and Gesu, L. Di and Lalla, N. Di and Donnarumma, I. and Dov{\v c}iak, M. and Ehlert, S. R. and Enoto, T. and Evangelista, Y. and Fabiani, S. and Ferrazzoli, R. and Garcia, J. A. and Gunji, S. and Hayashida, K. and Iwakiri, W. and Jorstad, S. G. and Kaaret, P. and Karas, V. and Kislat, F. and Kitaguchi, T. and Kolodziejczak, J. J. and Krawczynski, H. and Latronico, L. and Liodakis, I. and Maldera, S. and Manfreda, A. and Marin, F. and Marscher, A. P. and Marshall, H. L. and Massaro, F. and Matt, G. and Mitsuishi, I. and Mizuno, T. and Muleri, F. and Negro, M. and Ng, C. -Y. and O'Dell, S. L. and Omodei, N. and Oppedisano, C. and Pavlov, G. G. and Peirson, A. L. and Perri, M. and {Pesce-Rollins}, M. and Petrucci, P. -O. and Pilia, M. and Possenti, A. and Puccetti, S. and Ramsey, B. D. and Rankin, J. and Ratheesh, A. and Roberts, O. J. and Romani, R. W. and Sgr{\`o}, C. and Slane, P. and Soffitta, P. and Spandre, G. and Swartz, D. A. and Tamagawa, T. and Tavecchio, F. and Taverna, R. and Tawara, Y. and Tennant, A. F. and Thomas, N. E. and Tombesi, F. and Trois, A. and Turolla, R. and Vink, J. and Weisskopf, M. C. and Wu, K. and Xie, F. and Zane, S.},
  year = {2023},
  month = sep,
  journal = {MNRAS},
  volume = {524},
  pages = {2004--2014},
  publisher = {OUP},
  doi = {10.1093/mnras/stad1961}
}

@article{nagel1981,
  title = {Radiative {{Transfer}} in a {{Strongly Magnetized Plasma}} - {{Part Two}} - {{Effects}} of {{Comptonization}}},
  author = {Nagel, W.},
  year = {1981},
  month = dec,
  journal = {ApJ},
  volume = {251},
  pages = {288},
  publisher = {IOP},
  doi = {10.1086/159464}
}

@article{nasaheasarc2014,
  title = {{{HEAsoft}}: {{Unified Release}} of {{FTOOLS}} and {{XANADU}}},
  shorttitle = {{{HEAsoft}}},
  author = {{NASA HEASARC}},
  year = {2014},
  month = aug,
  journal = {ASCL},
  volume = {1408},
  pages = {004}
}

@article{nelson1993,
  title = {Nonthermal {{Cyclotron Emission}} from {{Low-Luminosity Accretion}} onto {{Magnetic Neutron Stars}}},
  author = {Nelson, Robert W. and Salpeter, E. E. and Wasserman, Ira},
  year = {1993},
  month = dec,
  journal = {ApJ},
  volume = {418},
  pages = {874},
  publisher = {IOP},
  doi = {10.1086/173445}
}

@article{nelson1995,
  title = {A {{Potential Cyclotron Line Signature}} in {{Low-Luminosity X-Ray Sources}}},
  author = {Nelson, Robert W. and Wang, John C. L. and Salpeter, E. E. and Wasserman, Ira},
  year = {1995},
  month = jan,
  journal = {ApJ},
  volume = {438},
  pages = {L99},
  publisher = {IOP},
  doi = {10.1086/187725}
}

@article{olausen2014,
  title = {The {{McGill Magnetar Catalog}}},
  author = {Olausen, S. A. and Kaspi, V. M.},
  year = {2014},
  month = may,
  journal = {ApJS},
  volume = {212},
  number = {1},
  pages = {6},
  doi = {10.1088/0067-0049/212/1/6},
  langid = {english}
}

@inproceedings{orlandini2001,
  title = {Hard {{X-ray}} Tails and Cyclotron Features in {{X-ray}} Pulsars},
  booktitle = {X-{{RAY ASTRONOMY}}: {{Stellar Endpoints}}, {{AGN}}, and the {{Diffuse X-ray Background}}.},
  author = {Orlandini, Mauro and Fiume, Daniele Dal},
  editor = {White, Nicholas E. and Malaguti, Guiseppe and Palumbo, Giorgio G.C.},
  year = {2001},
  month = dec,
  series = {{{AIP Conference Proceedings}}},
  volume = {599},
  pages = {283--294},
  publisher = {American Institute of Physics},
  address = {Melville, NY},
  doi = {10.1063/1.1434642}
}

@article{pike2023,
  title = {Accretion {{Spin-up}} and a {{Strong Magnetic Field}} in the {{Slow-spinning Be X-Ray Binary MAXI J0655-013}}},
  author = {Pike, Sean N. and Sugizaki, Mutsumi and van den Eijnden, Jakob and Coughenour, Benjamin and Jaodand, Amruta D. and Mihara, Tatehiro and Motta, Sara E. and Negoro, Hitoshi and Shaw, Aarran W. and Shidatsu, Megumi and Tomsick, John A.},
  year = {2023},
  month = sep,
  journal = {ApJ},
  volume = {954},
  pages = {48},
  publisher = {IOP},
  doi = {10.3847/1538-4357/ace696}
}

@article{pradhan2021,
  title = {Comprehensive Broad-Band Study of Accreting Neutron Stars with {{Suzaku}}: {{Is}} There a Bi-Modality in the {{X-ray}} Spectrum?},
  author = {Pradhan, Pragati and Paul, Biswajit and Bozzo, Enrico and Maitra, Chandreyee and Paul, B C},
  year = {2021},
  month = jan,
  journal = {MNRAS},
  volume = {502},
  number = {1},
  eprint = {https://academic.oup.com/mnras/article-pdf/502/1/1163/36171547/stab024.pdf},
  pages = {1163--1190},
  doi = {10.1093/mnras/stab024}
}

@article{predehl2021,
  title = {The {{eROSITA X-ray}} Telescope on {{SRG}}},
  author = {Predehl, P. and Andritschke, R. and Arefiev, V. and Babyshkin, V. and Batanov, O. and Becker, W. and B{\"o}hringer, H. and Bogomolov, A. and Boller, T. and Borm, K. and Bornemann, W. and Br{\"a}uninger, H. and Br{\"u}ggen, M. and Brunner, H. and Brusa, M. and Bulbul, E. and Buntov, M. and Burwitz, V. and Burkert, W. and Clerc, N. and Churazov, E. and Coutinho, D. and Dauser, T. and Dennerl, K. and Doroshenko, V. and Eder, J. and Emberger, V. and Eraerds, T. and Finoguenov, A. and Freyberg, M. and Friedrich, P. and Friedrich, S. and F{\"u}rmetz, M. and Georgakakis, A. and Gilfanov, M. and Granato, S. and Grossberger, C. and Gueguen, A. and Gureev, P. and Haberl, F. and H{\"a}lker, O. and Hartner, G. and Hasinger, G. and Huber, H. and Ji, L. and v. Kienlin, A. and Kink, W. and Korotkov, F. and Kreykenbohm, I. and Lamer, G. and Lomakin, I. and Lapshov, I. and Liu, T. and Maitra, C. and Meidinger, N. and Menz, B. and Merloni, A. and Mernik, T. and Mican, B. and Mohr, J. and M{\"u}ller, S. and Nandra, K. and Nazarov, V. and Pacaud, F. and Pavlinsky, M. and Perinati, E. and Pfeffermann, E. and Pietschner, D. and {Ramos-Ceja}, M. E. and Rau, A. and Reiffers, J. and Reiprich, T. H. and Robrade, J. and Salvato, M. and Sanders, J. and Santangelo, A. and Sasaki, M. and Scheuerle, H. and Schmid, C. and Schmitt, J. and Schwope, A. and Shirshakov, A. and Steinmetz, M. and Stewart, I. and Str{\"u}der, L. and Sunyaev, R. and Tenzer, C. and Tiedemann, L. and Tr{\"u}mper, J. and Voron, V. and Weber, P. and Wilms, J. and Yaroshenko, V.},
  year = {2021},
  month = mar,
  journal = {A\&A},
  volume = {647},
  pages = {A1},
  doi = {10.1051/0004-6361/202039313}
}

@article{rai2023,
  title = {Study of Recently Discovered {{Be}}/{{X-ray}} Pulsar {{MAXI J0655-013}} Using {{NuSTAR}}},
  author = {Rai, Binay and Tobrej, Mohammed and Ghising, Manoj and Tamang, Ruchi and Paul, Bikash Chandra},
  year = {2023},
  month = jul,
  journal = {MNRAS},
  volume = {524},
  number = {1},
  eprint = {2306.14970},
  primaryclass = {astro-ph},
  pages = {1352--1359},
  doi = {10.1093/mnras/stad1944},
  archiveprefix = {arXiv}
}

@article{rai2025,
  title = {{{NuSTAR}} and {{NICER}} Observations of {{X Persei}}},
  author = {Rai, Binay and Tobrej, Mohammed and Ghising, Manoj and Paul, Bikash Chandra},
  year = {2025},
  month = mar,
  journal = {JHEAP},
  volume = {45},
  pages = {265--272},
  publisher = {Elsevier},
  doi = {10.1016/j.jheap.2024.12.009}
}

@article{raman2023,
  title = {Quiet, but Not Silent: Uncovering Quiescent State Properties of Two Transient High-Mass {{X-ray}} Binaries},
  shorttitle = {Quiet, but Not Silent},
  author = {Raman, Gayathri and {Varun} and Pradhan, Pragati and Kennea, Jamie},
  year = {2023},
  month = dec,
  journal = {MNRAS},
  volume = {526},
  pages = {3262--3272},
  publisher = {OUP},
  doi = {10.1093/mnras/stad2577}
}

@article{rodi2024,
  title = {X {{Persei}}: {{A}} Study on the Origin of Its High-Energy Emission},
  shorttitle = {X {{Persei}}},
  author = {Rodi, J. and Natalucci, L. and Fiocchi, M.},
  year = {2024},
  month = jun,
  journal = {A\&A},
  volume = {689},
  pages = {A186},
  doi = {10.48550/arXiv.2406.15018}
}

@article{roucoescorial2018,
  title = {Discovery of Accretion-Driven Pulsations in the Prolonged Low {{X-ray}} Luminosity State of the {{Be}}/{{X-ray}} Transient {{GX}} 304-1},
  author = {Rouco Escorial, A. and {van den Eijnden}, J. and Wijnands, R.},
  year = {2018},
  month = dec,
  journal = {A\&A},
  volume = {620},
  pages = {L13},
  doi = {10.1051/0004-6361/201834572}
}

@article{roucoescorial2020,
  title = {Recurrent Low-Level Luminosity Behaviour after a Giant Outburst in the {{Be}}/{{X-ray}} Transient {{4U}} 0115+63},
  author = {Rouco Escorial, A. and Wijnands, R. and {van den Eijnden}, J. and Patruno, A. and Degenaar, N. and Parikh, A. and Ootes, L. S.},
  year = {2020},
  month = jun,
  journal = {A\&A},
  volume = {638},
  pages = {A152},
  publisher = {EDP},
  doi = {10.1051/0004-6361/201936287}
}

@article{salganik2023,
  title = {{{RX J0440}}.9+4431: Another Supercritical {{X-ray}} Pulsar},
  shorttitle = {{{RX J0440}}.9+4431},
  author = {Salganik, Alexander and Tsygankov, Sergey S. and Doroshenko, Victor and Molkov, Sergey V. and Lutovinov, Alexander A. and Mushtukov, Alexander A. and Poutanen, Juri},
  year = {2023},
  month = oct,
  journal = {MNRAS},
  volume = {524},
  number = {4},
  pages = {5213--5224},
  doi = {10.1093/mnras/stad2124},
  langid = {english}
}

@article{salganik2025,
  title = {Discovery of a Bimodal Luminosity Distribution in the Persistent {{Be}}/{{X-ray}} Pulsar {{2RXP J130159}}.6--635806},
  author = {Salganik, Alexander and Tsygankov, Sergey S. and Chernyakova, Maria and Malyshev, Denys and Poutanen, Juri},
  year = {2025},
  month = jun,
  journal = {A\&A},
  volume = {698},
  pages = {A71},
  doi = {10.1051/0004-6361/202453378},
  langid = {english}
}

@article{sartore2015,
  title = {The {{INTEGRAL}}/{{SPI}} View of {{A0535}}+26 during the {{Giant Outburst}} of 2011 {{February}}},
  author = {Sartore, N. and Jourdain, E. and Roques, J. P.},
  year = {2015},
  month = jun,
  journal = {ApJ},
  volume = {806},
  pages = {193},
  publisher = {IOP},
  doi = {10.1088/0004-637X/806/2/193}
}

@article{sokolova-lapa2021,
  title = {X-Ray Emission from Magnetized Neutron Star Atmospheres at Low Mass-Accretion Rates. {{I}}. {{Phase-averaged}} Spectrum},
  author = {{Sokolova-Lapa}, E. and Gornostaev, M. and Wilms, J. and Ballhausen, R. and Falkner, S. and Postnov, K. and Thalhammer, P. and F{\"u}rst, F. and Garc{\'i}a, J. A. and Shakura, N. and Becker, P. A. and Wolff, M. T. and Pottschmidt, K. and H{\"a}rer, L. and Malacaria, C.},
  year = {2021},
  month = jul,
  journal = {A\&A},
  volume = {651},
  pages = {A12},
  doi = {10.1051/0004-6361/202040228}
}

@article{sokolova-lapa2023a,
  title = {Low-Luminosity Accretion in {{Cep X-4}} during the Transition to Quiescence Observed by {{NuSTAR}}},
  author = {{Sokolova-Lapa}, E. and Zalot, N. and Stierhof, J. and Zainab, A. and Wilms, J. and Ballhausen, R. and Malacaria, C. and Kretschmar, P. and Escorial, A. Rouco and Pottschmidt, K. and Fuerst, F. and Ferrigno, Carlo and Pradhan, P. and Coley, Joel B.},
  year = {2023},
  month = aug,
  journal = {ATel},
  volume = {16171}
}

@article{sokolova-lapa2023b,
  title = {Vacuum Polarization Alters the Spectra of Accreting {{X-ray}} Pulsars},
  author = {{Sokolova-Lapa}, E. and Stierhof, J. and Dauser, T. and Wilms, J.},
  year = {2023},
  month = jun,
  journal = {A\&A},
  volume = {674},
  pages = {L2},
  doi = {10.1051/0004-6361/202346265}
}

@article{staubert2019,
  title = {Cyclotron Lines in Highly Magnetized Neutron Stars},
  author = {Staubert, R. and Tr{\"u}mper, J. and Kendziorra, E. and Klochkov, D. and Postnov, K. and Kretschmar, P. and Pottschmidt, K. and Haberl, F. and Rothschild, R. E. and Santangelo, A. and Wilms, J. and Kreykenbohm, I. and F{\"u}rst, F.},
  year = {2019},
  month = feb,
  journal = {A\&A},
  volume = {622},
  pages = {A61},
  doi = {10.1051/0004-6361/201834479}
}

@article{struder2001,
  title = {The {{European Photon Imaging Camera}} on {{XMM-Newton}}: {{The}} Pn-{{CCD}} Camera},
  shorttitle = {The {{European Photon Imaging Camera}} on {{XMM-Newton}}},
  author = {Str{\"u}der, L. and Briel, U. and Dennerl, K. and Hartmann, R. and Kendziorra, E. and Meidinger, N. and Pfeffermann, E. and Reppin, C. and Aschenbach, B. and Bornemann, W. and Br{\"a}uninger, H. and Burkert, W. and Elender, M. and Freyberg, M. and Haberl, F. and Hartner, G. and Heuschmann, F. and Hippmann, H. and Kastelic, E. and Kemmer, S. and Kettenring, G. and Kink, W. and Krause, N. and M{\"u}ller, S. and Oppitz, A. and Pietsch, W. and Popp, M. and Predehl, P. and Read, A. and Stephan, K. H. and St{\"o}tter, D. and Tr{\"u}mper, J. and Holl, P. and Kemmer, J. and Soltau, H. and St{\"o}tter, R. and Weber, U. and Weichert, U. and {von Zanthier}, C. and Carathanassis, D. and Lutz, G. and Richter, R. H. and Solc, P. and B{\"o}ttcher, H. and Kuster, M. and Staubert, R. and Abbey, A. and Holland, A. and Turner, M. and Balasini, M. and Bignami, G. F. and La Palombara, N. and Villa, G. and Buttler, W. and Gianini, F. and Lain{\'e}, R. and Lumb, D. and Dhez, P.},
  year = {2001},
  month = jan,
  journal = {A\&A},
  volume = {365},
  pages = {L18-L26},
  doi = {10.1051/0004-6361:20000066}
}

@article{sugizaki2015,
  title = {Luminosity and Spin-Period Evolution of {{GX}} 304-1 during Outbursts from 2009 to 2013 Observed with the {{MAXI}}/{{GSC}}, {{RXTE}}/{{PCA}}, and {{Fermi}}/{{GBM}}},
  author = {Sugizaki, Mutsumi and Yamamoto, Takayuki and Mihara, Tatehiro and Nakajima, Motoki and Makishima, Kazuo},
  year = {2015},
  month = aug,
  journal = {PASJ},
  volume = {67},
  number = {4},
  pages = {73},
  doi = {10.1093/pasj/psv039},
  langid = {english}
}

@article{thompson1995,
  title = {The Soft Gamma Repeaters as Very Strongly Magnetized Neutron Stars - {{I}}. {{Radiative}} Mechanism for Outbursts},
  author = {Thompson, Christopher and Duncan, Robert C.},
  year = {1995},
  month = jul,
  journal = {MNRAS},
  volume = {275},
  number = {2},
  pages = {255--300},
  doi = {10.1093/mnras/275.2.255},
  langid = {english}
}

@article{titarchuk1994,
  title = {Generalized {{Comptonization Models}} and {{Application}} to the {{Recent High-Energy Observations}}},
  author = {Titarchuk, Lev},
  year = {1994},
  month = oct,
  journal = {ApJ},
  volume = {434},
  pages = {570},
  publisher = {IOP},
  doi = {10.1086/174760}
}

@article{tsygankov2019,
  title = {Dramatic Spectral Transition of {{X-ray}} Pulsar {{GX}} 304-1 in Low Luminous State},
  author = {Tsygankov, S. S. and Rouco Escorial, A. and Suleimanov, V. F. and Mushtukov, A. A. and Doroshenko, V. and Lutovinov, A. A. and Wijnands, R. and Poutanen, J.},
  year = {2019},
  month = feb,
  journal = {MNRAS},
  volume = {483},
  eprint = {1810.13307},
  primaryclass = {astro-ph.HE},
  pages = {L144-L148},
  doi = {10.1093/mnrasl/sly236},
  archiveprefix = {arXiv}
}

@article{tsygankov2019a,
  title = {Cyclotron Emission, Absorption, and the Two Faces of {{X-ray}} Pulsar {{A}} 0535+262},
  author = {Tsygankov, Sergey S. and Doroshenko, Victor and Mushtukov, A. A. and Lutovinov, Alexander A. and Poutanen, Juri},
  year = {2019},
  month = jul,
  journal = {MNRAS},
  volume = {487},
  number = {1},
  eprint = {1905.09496},
  primaryclass = {astro-ph.HE},
  pages = {L30-L34},
  doi = {10.1093/mnrasl/slz079},
  archiveprefix = {arXiv}
}

@article{turner2001,
  title = {The {{European Photon Imaging Camera}} on {{XMM-Newton}}: {{The MOS}} Cameras},
  shorttitle = {The {{European Photon Imaging Camera}} on {{XMM-Newton}}},
  author = {Turner, M. J. L. and Abbey, A. and Arnaud, M. and Balasini, M. and Barbera, M. and Belsole, E. and Bennie, P. J. and Bernard, J. P. and Bignami, G. F. and Boer, M. and Briel, U. and Butler, I. and Cara, C. and Chabaud, C. and Cole, R. and Collura, A. and Conte, M. and Cros, A. and Denby, M. and Dhez, P. and Di Coco, G. and Dowson, J. and Ferrando, P. and Ghizzardi, S. and Gianotti, F. and Goodall, C. V. and Gretton, L. and Griffiths, R. G. and Hainaut, O. and Hochedez, J. F. and Holland, A. D. and Jourdain, E. and Kendziorra, E. and Lagostina, A. and Laine, R. and La Palombara, N. and Lortholary, M. and Lumb, D. and Marty, P. and Molendi, S. and Pigot, C. and Poindron, E. and Pounds, K. A. and Reeves, J. N. and Reppin, C. and Rothenflug, R. and Salvetat, P. and Sauvageot, J. L. and Schmitt, D. and Sembay, S. and Short, A. D. T. and Spragg, J. and Stephen, J. and Str{\"u}der, L. and Tiengo, A. and Trifoglio, M. and Tr{\"u}mper, J. and Vercellone, S. and Vigroux, L. and Villa, G. and Ward, M. J. and Whitehead, S. and Zonca, E.},
  year = {2001},
  month = jan,
  journal = {A\&A},
  volume = {365},
  pages = {L27-L35},
  doi = {10.1051/0004-6361:20000087}
}

@article{verner1996,
  title = {Atomic {{Data}} for {{Astrophysics}}. {{II}}. {{New Analytic Fits}} for {{Photoionization Cross Sections}} of {{Atoms}} and {{Ions}}},
  author = {Verner, D. A. and Ferland, G. J. and Korista, K. T. and Yakovlev, D. G.},
  year = {1996},
  month = jul,
  journal = {ApJ},
  volume = {465},
  pages = {487},
  doi = {10.1086/177435},
  langid = {english}
}

@article{vybornov2017,
  title = {Luminosity-Dependent Changes of the Cyclotron Line Energy and Spectral Hardness in {{Cepheus X-4}}},
  author = {Vybornov, V. and Klochkov, D. and Gornostaev, M. and Postnov, K. and {Sokolova-Lapa}, E. and Staubert, R. and Pottschmidt, K. and Santangelo, A.},
  year = {2017},
  month = may,
  journal = {A\&A},
  volume = {601},
  number = {A126},
  eprint = {1702.06361},
  primaryclass = {astro-ph.HE},
  pages = {A126},
  doi = {10.1051/0004-6361/201630275},
  archiveprefix = {arXiv}
}

@inproceedings{weisskopf2000,
  title = {Chandra {{X-ray Observatory}} ({{CXO}}): Overview},
  shorttitle = {Chandra {{X-ray Observatory}} ({{CXO}})},
  booktitle = {Proceedings {{Volume X-Ray Optics}}, {{Instruments}}, and {{Missions III}}, (2000)},
  author = {Weisskopf, Martin C. and Tananbaum, Harvey D. and Van Speybroeck, Leon P. and O'Dell, Stephen L.},
  editor = {Truemper, Joachim E. and Aschenbach, Bern},
  year = {2000},
  month = jul,
  series = {Proceedings of {{SPIE}}},
  volume = {4012},
  pages = {2--16},
  publisher = {SPIE},
  doi = {10.1117/12.391545}
}

@article{white1976a,
  title = {The {{X-ray}} Behaviour of {{3U}} 0352+30 ({{X Per}}).},
  author = {White, N. E. and Mason, K. O. and Sanford, P. W. and Murdin, P.},
  year = {1976},
  month = jul,
  journal = {MNRAS},
  volume = {176},
  pages = {201--215},
  publisher = {OUP},
  doi = {10.1093/mnras/176.1.201}
}

@article{wilms2000,
  title = {On the {{Absorption}} of {{X-rays}} in the {{Interstellar Medium}}},
  author = {Wilms, J. and Allen, A. and McCray, R.},
  year = {2000},
  month = oct,
  journal = {ApJ},
  volume = {542},
  number = {2},
  eprint = {astro-ph/0008425},
  pages = {914--924},
  doi = {10.1086/317016},
  archiveprefix = {arXiv}
}

@article{yamamoto2011,
  title = {Discovery of a {{Cyclotron Resonance Feature}} in the {{X-Ray Spectrum}} of {{GX}} 304-1 with {{RXTE}} and {{Suzaku}} during {{Outbursts Detected}} by {{MAXI}} in 2010},
  author = {Yamamoto, Takayuki and Sugizaki, Mutsumi and Mihara, Tatehiro and Nakajima, Motoki and Yamaoka, Kazutaka and Matsuoka, Masaru and Morii, Mikio and Makishima, Kazuo},
  year = {2011},
  month = nov,
  journal = {PASJ},
  volume = {63},
  pages = {S751-S757},
  publisher = {OUP},
  doi = {10.1093/pasj/63.sp3.S751}
}

@article{yatabe2018,
  title = {An Application of the {{Ghosh}} \& {{Lamb}} Model to the Accretion-Powered {{X-ray}} Pulsar {{X Persei}}},
  author = {Yatabe, Fumiaki and Makishima, Kazuo and Mihara, Tatehiro and Nakajima, Motoki and Sugizaki, Mutsumi and Kitamoto, Shunji and Yoshida, Yuki and Takagi, Toshihiro},
  year = {2018},
  month = oct,
  journal = {PASJ},
  volume = {70},
  pages = {89},
  publisher = {OUP},
  doi = {10.1093/pasj/psy088}
}

@article{zalot2024,
  title = {An In-Depth Analysis of the Variable Cyclotron Lines in {{GX}} 301-2},
  author = {Zalot, Nicolas and {Sokolova-Lapa}, Ekaterina and Stierhof, Jakob and Ballhausen, Ralf and Zainab, Aafia and Pottschmidt, Katja and F{\"u}rst, Felix and Thalhammer, Philipp and Islam, Nazma and Diez, Camille M. and Kretschmar, Peter and Berger, Katrin and Rothschild, Richard and Malacaria, Christian and Pradhan, Pragati and Wilms, J{\"o}rn},
  year = {2024},
  month = jun,
  journal = {A\&A},
  volume = {686},
  pages = {A95},
  doi = {10.1051/0004-6361/202348841},
  copyright = {https://creativecommons.org/licenses/by/4.0}
}

@article{zeldovich1969,
  title = {X-{{Ray Emission Accompanying}} the {{Accretion}} of {{Gas}} by a {{Neutron Star}}},
  author = {Zel'dovich, {\relax Ya}. B. and Shakura, N. I.},
  year = {1969},
  month = oct,
  journal = {SvA},
  volume = {13},
  pages = {175}
}

\begin{appendix}

\section{Data extraction and reduction}
In this section we document our data extraction procedure.
All extractions are conducted with HEAsoft version 6.36 \citep{nasaheasarc2014}, all subsequent spectral analysis is conducted with \texttt{ISIS} version 1.6.2 \citep{houck2000}.
All spectra are rebinned based on a minimum counts per bin criterion, whose threshold value is set manually for each source and instrument depending on the count rate.

\subsection{\nustar}
\nustar datasets are reprocessed using \nustar pipeline version 0.4.12.
For each focal plane module, source and background regions are chosen as circles and annuli, respectively.
Both the source and background regions are centered on the point source. 
As the observations cover a wide range of source fluxes, the source region circle radii are chosen individually for each observation. 
The background region annuli radii are chosen such that they cover as much of the detector area as possible in order to constrain the background emission as precisely as possible.
A potential drawback of this method is that the detector response is not uniform over the full detector area.
For \srga, we instead chose four circular regions as the background region, as the observation was significantly affected by stray light \citep{ludlam2022}.

We extract X-ray spectra using NuSTARDAS version 2.1.5.
Unless given otherwise in Sect.~\ref{sec:spectral_analysis}, we make use of the full \nustar energy range of 3.5--79\,keV.

\subsection{\chandra}
\chandra data are extracted using \texttt{CIAO} version~4.17. The source and background regions are circular regions. 
\subsection{\nicer}
\textit{NICER/XTI} data are extracted using the \texttt{nicerl3} version~1.14 routines. We utilize the SCORPEON background model \citep{markwardt2024}.
\subsection{\swift}
\xrt data are extracted using the \texttt{xrtpipeline} version 0.13.7. Source and background regions are chosen as circles and annuli around the point sources, respectively.
\subsection{\xmm}
\xmm data are extracted using the Science Analysis System (SAS) version 22.1.0. Source and background regions are circular, where the source region radius depends on the source flux and the background source region is set to 60".
Flaring particle background was taken into account by visual inspection of the 10--12\,keV light curve; time intervals with visible spikes in count rate have been discarded. Additionally, a significant portion of the \xmm observation of \maxiZero had to be discarded to to solar activity.

\onecolumn
\begin{landscape}
\section{Source overview and spectral parameters}
\renewcommand{\arraystretch}{1.35}
\begin{longtable}{lcccccc}
	\caption{Overview of observations and spectral models utilized in this work. Upper part of the table: source and observation information. Lower part of the table: parameters of \texttt{doublehump} spectral fits shown in Fig.~\ref{fig:doublehump}.}
	\\\endfirsthead
	\\\hline

	& \rotatebox[origin=c]{0}{X~Per} & \rotatebox[origin=c]{0}{GRO~J1008$-$57} & \rotatebox[origin=c]{0}{IGR~J21347$+$4737} & \rotatebox[origin=c]{0}{\begin{tabular}{c}2SXPS\\J075542.5$-$293353\end{tabular}} & \rotatebox[origin=c]{0}{GX~304$-$1} \\ \hline
	Distance\,[kpc]\tablefootmark{(a)} & $0.600\pm0.014$ & $3.52^{+0.17}_{-0.16}$ & $8.5^{+1.1}_{-0.8}$ & $3.29^{+0.12}_{-0.14}$ & $1.85^{+0.06}_{-0.05}$ \\
	$P_\mathrm{orb}\,[\mathrm{d}]$ & $250.3\pm0.6$ / \citetalias{delgado-marti2001} & $249.48\pm0.04$ / \citetalias{kuhnel2013} & ${\sim}34.3$ / \citetalias{gorban2022a} & ${\sim}260$ / \citetalias{doroshenko2021} & $132.189\pm0.02$ / \citetalias{sugizaki2015} \\
	$P_\mathrm{pulse}\,[\mathrm{s}]$ & $833.5\pm0.2$ / \citetalias{rai2025} & $\eqsim 93.2$ / \citetalias{lutovinov2021} & $322.738\pm0.018$ / \citetalias{ghising2023a} & $308.26\pm0.02$ / \citetalias{doroshenko2021} & $274.9817 \pm 0.0001$ / \citetalias{malacaria2015} \\
	\nustar OBS ID & 30401033002 & 90501357002 & 90601339002 & \begin{tabular}{c}80102103002\\90601322001\end{tabular} & 30701015002 \\
	\nustar date & 2019-01-01 & 2020-01-01 & 2020-12-17 & \begin{tabular}{c}2016-03-26\\2020-07-19\end{tabular} & 2022-01-29 \\
	Joint with & -- & \swift{} & XMM\tablefootmark{(b)} & \chandra{} & \textit{XMM}\tablefootmark{(c)} \\
	\nustar exp.\,[ks] / PI & 58.3 / M. Wolff & 45.2 / A. Lutovinov (TOO) & 27.1 / S. Pike (TOO) & \begin{tabular}{c}54.7 / V. Kaspi\\49.8 / G. Younes\end{tabular} & 174.0 / E. Sokolova-Lapa \\ \hline
	$C_\mathrm{FPMB}$ & $0.980\pm0.003$ & $1.02\pm0.04$ & $1.02\pm0.03$ & $1.01\pm0.03$ & $1.04\pm0.02$ \\
	$C_\mathrm{XRT/EPN/FPMA2}$ & -- & $0.8\pm0.3$ & $0.213^{+0.007}_{-0.006}$ & $0.88\pm0.03$ & $0.84\pm0.04$ \\
	$C_\mathrm{EMOS1/FPMB2}$ & -- & -- & $0.180\pm0.007$ & $0.90\pm0.03$ & $0.72\pm0.04$ \\
	$C_\mathrm{EMOS2/ACIS}$ & -- & -- & $0.206\pm0.008$ & $0.24\pm0.02$ & $0.85\pm0.04$ \\
	$N_\mathrm{H}\,[10^{22}\,\mathrm{cm}^{-2}]$ & $3.3\pm0.4$ & $3\pm2$ & $0.48\pm0.09$ & $0.8\pm0.3$ & $1.7\pm0.3$ \\
	$N_\texttt{cutoffpl}\left[\times 10^{-3}\right]$ & $180\pm9$ & $2.2^{+1.0}_{-0.7}$ & $0.77^{+0.08}_{-0.07}$ & $0.8\pm0.2$ & $1.8^{+0.4}_{-0.3}$ \\
	$\Gamma_1$ & $1.43\pm0.04$ & $1.4\pm0.4$ & $-0.3\pm0.2$ & $0.2\pm0.3$ & $0.1\pm0.2$ \\
	$E_\mathrm{fold,1}\,[\mathrm{keV}]$ & $10.1\pm0.5$ & $9^{+5}_{-3}$ & $3.4^{+0.5}_{-0.4}$ & $3.0\pm0.4$ & $2.2\pm0.2$ \\
	$E_\mathrm{fold,2}\,[\mathrm{keV}]$ & $15.2\pm0.6$ & $\left(3^{+7}_{-2}\right)\times10^{1}$ & $6.3^{+0.6}_{-0.5}$ & $8.7\pm0.6$ & $7.2\pm0.5$ \\
	$E_\mathrm{int}$\,[keV] & $28.1^{+0.7}_{-0.6}$ & $25\pm5$ & $14\pm2$ & $13.2\pm0.6$ & $12.8\pm0.4$ \\\hline
	$\mathcal{L}\,\left[10^{34}\,\mathrm{erg}\,\mathrm{s}^{-1}\right]$ \tablefootmark{(d)}& 3.23 & 0.96 & 6.48 & 0.38 & 0.24 \\
	$\chi^2 / \mathrm{d.o.f.}$ & $671 / 522$ & $195 / 181$ & $261 / 213$ & $222 / 189$ & $280 / 258$ \\
	$\chi^2_\mathrm{red}$ & $1.29$ & $1.08$ & $1.23$ & $1.18$ & $1.08$ \\\hline $\Delta \chi^2_\mathrm{cutoffpl}$ & $2764$ & $35$ & $163$ & $285$ & $667$ \\\hline

	\pagebreak  
	\caption{continued.}\\
	\hline \\

	& \rotatebox[origin=c]{0}{SRGA~J124404.1-632232} & \rotatebox[origin=c]{0}{MAXI~0655$-$013} & \rotatebox[origin=c]{0}{A0535+26} & \rotatebox[origin=c]{0}{Cep~X-4} \\ \hline
	Distance\,[kpc]\tablefootmark{(a)} & $5.8^{+0.8}_{-0.0}$ & $3.44^{+0.26}_{-0.21}$ & $1.77^{+0.07}_{-0.06}$ & $7.2^{+0.8}_{-0.7}$ &  \\
	$P_\mathrm{orb}\,[\mathrm{d}]$ & $138\pm1$ / \citetalias{mroz2021} & $27.9\pm1.7$ / \citetalias{rai2023} & $110.3\pm0.3$ / \citetalias{finger1994} & $20.85$ / \citetalias{mcbride2007} &  \\
	$P_\mathrm{pulse}\,[\mathrm{s}]$ & $538.70\pm0.05$ / \citetalias{doroshenko2022} & $1129.09\pm0.04$ / \citetalias{pike2023} & $103.3890\pm 0.0009$ / \citetalias{ballhausen2017} & $66.3336\pm0.0002$ / \citetalias{vybornov2017} &  \\
	\nustar OBS ID & 80660301002 & 31001020002 & 90401370001 & 90901322002 &  \\
	\nustar date & 2021-01-30 & 2024-10-02 & 2018-12-26 & 2023-07-07 &  \\
	Joint with & \nicer{} & \xmm{} & \swift{} & \swift{} &  \\
	\nustar exp.\,[ks] / PI & 54.0 / V. Doroshenko (TOO) & 84.1 / C. Malacaria & 54.9 / S. Tsygankov (TOO) & 26.5 / R. Ballhausen (TOO) &  \\ \hline
	$C_\mathrm{FPMB}$ & $1.02\pm0.04$ & $1.05\pm0.04$ & $0.97\pm0.02$ & $1.01\pm0.04$ &  \\
	$C_\mathrm{XTI/EPN/XRT}$ & $1.02^{+0.10}_{-0.09}$ & $0.76\pm0.04$ & $0.8\pm0.2$ & $0.6\pm0.3$ &  \\
	$C_\mathrm{EMOS1}$ & -- & $0.67^{+0.04}_{-0.03}$ & -- & -- &  \\
	$C_\mathrm{EMOS2}$ & -- & $0.72\pm0.04$ & -- & -- &  \\
	$N_\mathrm{H}\,[10^{22}\,\mathrm{cm}^{-2}]$ & $1.4\pm0.4$ & $0.9\pm0.2$ & $0.2^{+0.8}_{-0.2}$ & 0.8\tablefootmark{(e)} &  \\
	$N_\texttt{cutoffpl}\left[\times 10^{-3}\right]$ & $0.6^{+0.3}_{-0.2}$ & $0.33\pm0.04$ & $1.9^{+0.7}_{-0.5}$ & $0.6^{+0.7}_{-0.4}$ &  \\
	$\Gamma_1$ & $0.1\pm0.6$ & $-0.6^{+0.2}_{-0.3}$ & $-0.6\pm0.4$ & $-1\pm2$ &  \\
	$E_\mathrm{fold,1}\,[\mathrm{keV}]$ & $2.9^{+1.5}_{-0.9}$ & $2.0\pm0.3$ & $2.1^{+0.3}_{-0.2}$ & $1.7^{+1.0}_{-0.5}$ &  \\
	$E_\mathrm{fold,2}\,[\mathrm{keV}]$ & $6^{+3}_{-2}$ & $5.4^{+1.2}_{-0.9}$ & $6.2\pm0.2$ & $5.3^{+0.5}_{-0.4}$ &  \\
	$E_\mathrm{int}$\,[keV] & $13^{+4}_{-3}$ & $12\pm2$ & $10.6\pm0.3$ & $7.9^{+0.7}_{-0.5}$ &  \\\hline
	$\mathcal{L}\,\left[10^{34}\,\mathrm{erg}\,\mathrm{s}^{-1}\right]$ \tablefootmark{(d)}& 2.84 & 0.40 & 1.41 & 6.06 &  \\
	$\chi^2 / \mathrm{d.o.f.}$ & $144 / 130$ & $201 / 183$ & $438 / 317$ & $127 / 149$ &  \\
	$\chi^2_\mathrm{red}$ & $1.10$ & $1.10$ & $1.38$ & $0.85$ &  \\\hline $\Delta \chi^2_\mathrm{cutoffpl}$ & $40$ & $172$ & $441$ & $78$ &  \\\hline

	\label{tab:app1}
\end{longtable}
\tablefoot{

(a) Photogeometric distances from \textit{Gaia} DR3 \citep{gaiacollaboration2016,gaiacollaboration2023,bailer-jones2021}.

(b) The \textsl{XMM-Newton} observation of \igr was not performed simultaneous to the \nustar observation, but at similar luminosity.

(c) The \textsl{XMM-Newton} observation of \gxTOF was not performed simultaneous to the \nustar observation, but at the same orbital phase and similar luminosity.

(d) Absorbed luminosity in the observer's frame in the 5--50\,keV energy band.

(e) Galactic absorption column density $N_\mathrm{H}$ from \citet{hi4picollaboration2016}.
}
\tablebib{
Ba17: \citet{ballhausen2017}, De01: \citet{delgado-marti2001}, Do21: \citet{doroshenko2021}, Do22: \citet{doroshenko2022}, Fi94: \citet{finger1994}, Gh23: \citet{ghising2023a}, Go22: \citet{gorban2022a}, Ku13: \citet{kuhnel2013}, Lu21: \citet{lutovinov2021}, Ma15: \citet{malacaria2015}, Mc07: \citet{mcbride2007}, Mr21: \citet{mroz2021}, Pi23: \citet{pike2023}, Ra23: \citet{rai2023}, Ra25: \citet{rai2025}, Su15: \citet{sugizaki2015}, Vy17: \citet{vybornov2017}
}
\end{landscape}
\end{appendix}
\end{document}